\documentclass[12pt,english]{article}
\usepackage{times}
\usepackage[T1]{fontenc}
\usepackage{geometry}
\usepackage{rotating}
\usepackage{graphicx}
\usepackage{setspace}
\usepackage{amssymb}

\makeatletter

\providecommand{\tabularnewline}{\\}

\usepackage{bm}

\usepackage{babel}
\makeatother
\begin{document}

\section*{\noindent Memory-Enhanced Dynamic Evolutionary Control of Reconfigurable
Intelligent Surfaces}

\noindent ~

\noindent \vfill

\noindent F. Zardi,$^{(1)(2)}$ G. Oliveri,$^{(1)(2)}$ \emph{Senior
Member, IEEE}, and A. Massa,$^{(1)(2)(3)(4)(5)}$ \emph{Fellow, IEEE}

\noindent \vfill

\noindent {\footnotesize ~}{\footnotesize \par}

\noindent {\footnotesize $^{(1)}$} \emph{\footnotesize ELEDIA Research
Center} {\footnotesize (}\emph{\footnotesize ELEDIA}{\footnotesize @}\emph{\footnotesize UniTN}
{\footnotesize - University of Trento)}{\footnotesize \par}

\noindent {\footnotesize DICAM - Department of Civil, Environmental,
and Mechanical Engineering}{\footnotesize \par}

\noindent {\footnotesize Via Mesiano 77, 38123 Trento - Italy}{\footnotesize \par}

\noindent \textit{\emph{\footnotesize E-mail:}} {\footnotesize \{}\emph{\footnotesize francesco.zardi,
giacomo.oliveri, andrea.massa}{\footnotesize \}@}\emph{\footnotesize unitn.it}{\footnotesize \par}

\noindent {\footnotesize Website:} \emph{\footnotesize www.eledia.org/eledia-unitn}{\footnotesize \par}

\noindent {\footnotesize ~}{\footnotesize \par}

\noindent {\footnotesize $^{(2)}$} \emph{\footnotesize CNIT - \char`\"{}University
of Trento\char`\"{} ELEDIA Research Unit }{\footnotesize \par}

\noindent {\footnotesize Via Sommarive 9, 38123 Trento - Italy}{\footnotesize \par}

\noindent {\footnotesize Website:} \emph{\footnotesize www.eledia.org/eledia-unitn}{\footnotesize \par}

\noindent {\small ~}{\small \par}

\noindent {\footnotesize $^{(3)}$} \emph{\footnotesize ELEDIA Research
Center} {\footnotesize (}\emph{\footnotesize ELEDIA}{\footnotesize @}\emph{\footnotesize UESTC}
{\footnotesize - UESTC)}{\footnotesize \par}

\noindent {\footnotesize School of Electronic Science and Engineering,
Chengdu 611731 - China}{\footnotesize \par}

\noindent \textit{\emph{\footnotesize E-mail:}} \emph{\footnotesize andrea.massa@uestc.edu.cn}{\footnotesize \par}

\noindent {\footnotesize Website:} \emph{\footnotesize www.eledia.org/eledia}{\footnotesize -}\emph{\footnotesize uestc}{\footnotesize \par}

\noindent {\footnotesize ~}{\footnotesize \par}

\noindent {\footnotesize $^{(4)}$} \emph{\footnotesize ELEDIA Research
Center} {\footnotesize (}\emph{\footnotesize ELEDIA@TSINGHUA} {\footnotesize -
Tsinghua University)}{\footnotesize \par}

\noindent {\footnotesize 30 Shuangqing Rd, 100084 Haidian, Beijing
- China}{\footnotesize \par}

\noindent {\footnotesize E-mail:} \emph{\footnotesize andrea.massa@tsinghua.edu.cn}{\footnotesize \par}

\noindent {\footnotesize Website:} \emph{\footnotesize www.eledia.org/eledia-tsinghua}{\footnotesize \par}

\noindent {\small ~}{\small \par}

\noindent {\small $^{(5)}$} {\footnotesize School of Electrical Engineering}{\footnotesize \par}

\noindent {\footnotesize Tel Aviv University, Tel Aviv 69978 - Israel}{\footnotesize \par}

\noindent \textit{\emph{\footnotesize E-mail:}} \emph{\footnotesize andrea.massa@eng.tau.ac.il}{\footnotesize \par}

\noindent {\footnotesize Website:} \emph{\footnotesize https://engineering.tau.ac.il/}{\footnotesize \par}

\noindent \vfill

\noindent \emph{This work has been submitted to the IEEE for possible
publication. Copyright may be transferred without notice, after which
this version may no longer be accessible}

\noindent ~

\newpage
\section*{Memory-Enhanced Dynamic Evolutionary Control of Reconfigurable Intelligent
Surfaces}

~

~

~

\begin{flushleft}F. Zardi, G. Oliveri, and A. Massa\end{flushleft}

\vfill

\begin{abstract}
\noindent An innovative evolutionary method for the dynamic control
of reconfigurable intelligent surfaces (\emph{RIS}s) is proposed.
It leverages, on the one hand, on the exploration capabilities of
evolutionary strategies and their effectiveness in dealing with large-scale
discrete optimization problems and, on the other hand, on the implementation
of memory-enhanced search mechanisms to exploit the time/space correlation
of communication environments. Without modifying the base station
(\emph{BS}) beamforming strategy and using an accurate description
of the meta-atom response to faithfully account for the micro-scale
\emph{EM} interactions, the \emph{RIS} control (\emph{RISC}) algorithm
maximizes the worst-case \emph{throughput} across all users without
requiring that the Green's partial matrices, from the \emph{BS} to
the \emph{RIS} and from the \emph{RIS} to the users, be (separately)
known/measured. Representative numerical examples are reported to
illustrate the features and to assess the potentialities of the proposed
approach for the \emph{RISC}.

\vfill
\end{abstract}
\noindent \textbf{Key words}: Reconfigurable Passive \emph{EM} Skins;
Smart Electromagnetic Environment; Next-Generation Communications;
Metamaterials; Metasurfaces; Evolutionary Optimization; Genetic Algorithm.

\newpage
\section{Introduction and Motivation\label{sec:Introduction}}

\noindent The Smart Electromagnetic Environment (\emph{SEME}) paradigm
is the core of a revolutionary approach for the implementation, the
deployment, and the operation of wireless communication systems of
next generations mobile networks \cite{Massa 2021}-\cite{Yang 2022}.
As a matter of fact, the fundamental concept of a \emph{SEME}, that
is the opportunistic exploitation of the environment to improve the
quality-of-service (\emph{QoS}) at the users of a wireless network,
will require a deep revisitation of current communication systems
from both the architectural and the operative viewpoints \cite{Massa 2021}-\cite{Yang 2022}.
To implement the \emph{SEME} vision, several technologies with different
levels of complexity have been proposed including integrated access
and back-haul (\emph{IAB}) nodes, smart repeaters (\emph{SR}s), and
different classes of electromagnetic skins (\emph{EMS}s) \cite{Yang 2022}-\cite{Oliveri 2023}.
Reconfigurable intelligent surfaces (\emph{RIS}s) are Reconfigurable
Passive \emph{EM} Skins (\emph{RP-EMS}s), featuring an interesting
trade-off between power consumption, implementation complexity, fabrication
costs, and flexibility/reconfigurability \cite{Yang 2022}\cite{Oliveri 2022d}.
A \emph{RIS} usually consists of a combination of elementary meta-atoms,
each one embedding at least one electronically-reconfigurable component
(e.g., a diode) that controls its scattering parameters \cite{Yang 2022}\cite{Oliveri 2022d}.
By individually adjusting the state of each meta-atom, the \emph{RIS}
modifies its local scattering properties so that the macro-scale reflected
field fulfils a desired property (e.g., an anomalous direction of
reflection) \cite{Yang 2022}\cite{Oliveri 2022d}. Accordingly, a
\emph{RIS} is a passive reflecting device (i.e., the scattered power
is always equal or lower than the incident one) that adaptively tailors
the wireless response of the environment \cite{Yang 2022}\cite{Oliveri 2022d}.
Therefore, there has been a considerable interest, from both the academic
and the industrial communities, on the design, the implementation,
and the validation of \emph{RIS}s in wireless communications \cite{Wu 2019}\cite{Yang 2022}\cite{Li 2022}-\cite{Zhang 2021a}.
Despite the successful results, several open challenges still exist
from the algorithmic and the technological viewpoints before a widespread
deployment of \emph{RIS}s in large-scale applications. For instance,
one of the most challenging items to be addressed is the definition
of a suitable \emph{RIS} control (\emph{RISC}) to be (\emph{i}) used
for dynamically adjusting the meta-atom status to yield the envisioned
performance improvements at the network level and (\emph{ii}) easily
integrated into the existing infrastructures in an almost seamseless
fashion with minimum efforts and costs.

\noindent It is worthwhile to remember that many interesting \emph{RISC}
approaches have been proposed in the literature that consider different
performance objectives (e.g., minimum transmitter power \cite{Wu 2019}\cite{Li 2022}\cite{Wu 2020},
maximum signal-to-interference-plus noise ratio {[}\emph{SINR}{]}
\cite{Xie 2021}, maximum sum-rate capacity \cite{Guo 2020}), and/or
different constraints (e.g., per-user \emph{SINR} \cite{Wu 2019},
total transmitted power \cite{Xie 2021}, per-user capacity \cite{Huang 2019}),
and/or various optimization techniques (e.g., gradient methods \cite{Xie 2021}\cite{Guo 2020},
convex programming \cite{Li 2022}, as well as random walk \cite{Zhang 2021a}).
However, despite intrinsic differences, some common features exist
among these methods on both the operation principles and the assumptions
including the following ones:

\begin{itemize}
\item \noindent the modification of the beamforming algorithm of the base
station (\emph{BS}) to account for the \emph{RIS} presence \cite{Wu 2019}\cite{Li 2022}-\cite{Zhang 2021a}.
This implies the replacement of the beamforming methods implemented
in current generation \emph{BS}s when adding \emph{RIS}s;
\item \noindent the knowledge of the Green's partial matrices from both
the \emph{BS} to the \emph{RIS} and the \emph{RIS} to the user terminals
\cite{Wu 2019}\cite{Li 2022}-\cite{Zhang 2021a}. Obtaining such
an information in practical scenarios is quite problematic since,
one the one hand, it requires the \emph{RIS} to include more advanced
receiving/decoding/processing capabilities, on the other hand, it
implies the exploitation of complex wireless channel estimation mechanisms
with a potentially cumbersome overhead \cite{Swindlehurst 2022};
\item \noindent the assumption that the \emph{RIS} meta-atom enables an
ideal control of the phase of the reflected wave over a continuous
range, sometimes of its magnitude \cite{Wu 2019}\cite{Li 2022}-\cite{Nadeem 2020},
as well. Nevertheless, standard unit cells permit a limited magnitude
adjustment and they generally feature very few bits to simplify the
control architecture and to reduce the overall costs \cite{Liang 2022}\cite{Oliveri 2022d};
\item \noindent the use of idealized \emph{EMS}s (e.g., point-like isotropic
scatterers), which take into account neither the actual physics of
the device nor the presence of induced surface currents \cite{Wu 2019}\cite{Li 2022}-\cite{Zhang 2021a},
for the validation of the \emph{RISC.} 
\end{itemize}
\noindent This work is aimed at proposing an innovative complement
to the existing \emph{RISC} approaches for a more easy and reliable
deployment of \emph{RIS}s in real scenarios. Therefore, the following
guidelines are fulfilled: (\emph{i}) the \emph{RISC} method is suitable
for existing wireless networks without requiring any modification
to the \emph{BS} beamforming algorithm; (\emph{ii}) only the overall
{}``cascaded'' \cite{Swindlehurst 2022} Green's matrix from the
\emph{BS} to the user terminals is available as it usually happens
in current multi-input multi-output (\emph{MIMO}) communication systems
\cite{Swindlehurst 2022}-\cite{Oliveri 2022e}, while there are no
information on the Green's partial matrices from/to the \emph{RIS}
(i.e., the \emph{RIS} is {}``blind'' with respect to the environment);
(\emph{iii}) the \emph{RISC} method accounts for the limited reconfigurability
of typical cost-effective implementations of the unit cells of the
\emph{RIS} \cite{Oliveri 2022d}\cite{Liang 2022} owing to the quantized
phase response. Accordingly, a dynamic discrete optimization is formulated
to maximize the minimum throughput across all users (i.e., \emph{max-min}
criterion to guarantee an adequate \emph{QoS} to all users) subject
to a fixed transmitted power and a noise level. The \emph{Generalized
Sheet Transition Condition} (\emph{GSTC}) approach \cite{Yang 2019}-\cite{Senior 1987},
whose accuracy/reliability has been already assessed in several \emph{SEME}
scenarios \cite{Oliveri 2022d}\cite{Oliveri 2021c}, is used to compute
the Green's matrix coefficients. Moreover, no direct line-of-sight
(\emph{LOS}) links between the \emph{BS} and the mobile terminals
are assumed (Fig. 1) in order to primarily analyze the \emph{RIS}
effects on the communication \emph{QoS}, while the zero-forcing (\emph{ZF})
method is adopted at the \emph{BS} \cite{Oliveri 2019e}\cite{Oliveri 2022e}
for the beamforming%
\footnote{\noindent The \emph{ZF} method \cite{Oliveri 2019e}\cite{Oliveri 2022e}
is very popular and diffused, but the same \emph{RISC} can be adopted
regardless of the beamforming method at the \emph{BS}.%
} .

\noindent Owing to the non-linear and discrete nature of the problem
at hand as well as the lack of knowledge on the {}``partial'' Green's
matrices, which prevents a separate optimization of the \emph{BS}
illumination and the \emph{RIS} configuration, an \emph{Evolutionary
Algorithm} (\emph{EA}) \cite{Caorsi 2003}-\cite{Benedetti 2008}
is chosen as solution tool by also exploiting the repetitiveness of
the communication scenario (e.g., the instantaneous locations of the
terminals) due to the self-similarity of the user trajectories. This
latter feature is of fundamental importance in the \emph{RISC} since
it allows a minimization of the communication overhead. In particular,
the memory-enhanced \emph{EA} (\emph{ME-EA}) method, previously proposed
for adaptive array control \cite{Caorsi 2003}\cite{Massa 2004},
is customized here to the \emph{RISC}.

\noindent To the best of the authors' knowledge, the main methodological
novelties of this work with respect to the state-of-the art literature
on the topic include 

\begin{itemize}
\item the introduction of a dynamic \emph{RISC} strategy that allows a considerable
reduction of both the complexity of the deployment of \emph{RIS}s
in current wireless network architectures and the communications overhead;
\item the adoption of a discrete optimization paradigm that, unlike widely
adopted methods \cite{Wu 2019}\cite{Li 2022}-\cite{Guo 2020}, accounts
for the limitations in the reconfiguration of non-ideal \emph{RIS}
meta-atoms;
\item the implementation of a memory-enhanced \emph{RISC} method that leverages
on the information gathered on the time-evolution of the propagation
environment rather than repeatedly re-initializing and optimizing
the \emph{RIS} status at each update of the communication scenario.
\end{itemize}
\noindent The outline of the paper is as follows. After the formulation
of the \emph{RISC} problem (Sect. \ref{sec:Problem-Formulation}),
the proposed \emph{RISC} method is described in Sect. \ref{sec:Control}.
A set of representative numerical results is then reported to assess
its effectiveness and reliability also in a comparative fashion (Sect.
\ref{sec:Results}). Finally, some conclusions and final remarks are
drawn in Section \ref{sec:Conclusions-and-Remarks}.

\section{\noindent Problem Formulation\label{sec:Problem-Formulation} }

\noindent With reference to a \emph{MIMO} scenario in a time-harmonic
regime, a \emph{BS} composed by $M$ single-polarization%
\footnote{\noindent The double-polarization case, which is omitted hereinafter
for notation simplicity, can be straightforwardly deduced within the
same framework.%
} radiators, the $m$-th ($m=1,...,M$) being located at $\mathbf{r}_{m}=\left(x_{m},y_{m},z_{m}\right)$,
and operating in the multi-user multi-antenna down-link communication
mode%
\footnote{\noindent For symmetry reasons, the up-link is neglected.%
} serves at the $c$-th ($c=1,...,C$) time instant $t^{\left(c\right)}$
the $L$ users in Fig. 1 whose positions are \{$\mathbf{r}_{l}^{\left(c\right)}$;
$l=1,...,L$\}. Since no direct wireless link can be established between
the \emph{BS} and these $L$ users because of the presence of buildings
(Fig. 1), the communication is then enabled by a \emph{RIS}, which
is controlled by the \emph{BS} through a dedicated digital channel
(Fig. 1). The \emph{RIS} $\Xi_{\mathrm{RIS}}$, which comprises $P$
meta-atoms, is installed on a supporting wall $\Xi_{\mathrm{wall}}$
and it is centered in the origin of the local coordinate systems $\left(x,y,z\right)=\left(0,0,0\right)$.
The system composed by the \emph{RIS} and the surrounding wall (Fig.
1) is illuminated at the $c$-th ($c=1,...,C$) time-instant by the
\emph{BS} with $B$ incident fields, the $b$-th ($b=1,...,B$) being
given by

\noindent \begin{equation}
\mathbf{E}_{b}^{\left(c\right)}\left(\mathbf{r}\right)=\sum_{m=1}^{M}\mathcal{G}\left(\mathbf{r}_{m},\mathbf{r}\right)\alpha_{m,b}^{\left(c\right)}\widehat{\bm{\chi}}\label{eq:}\end{equation}
where $\mathcal{G}$ is the Fourier transform, at the frequency $f_{0}$,
of the time-domain Green's tensor component modeling the electromagnetic
propagation between the input port of the $m$-th ($m=1,...,M$) transmitting
antenna and the vectorial field in $\mathbf{r}$ \cite{Oliveri 2019e}\cite{Oliveri 2022e},
$\widehat{\bm{\chi}}$ is the polarization unit vector of the \emph{BS}
antennas, and $\alpha_{m,b}^{\left(c\right)}$ is the $m$-th ($m=1,...,M$)
excitation of the \emph{BS} to afford the $b$-th ($b=1,...,B$) beam
at the $c$-th ($c=1,...,C$) time-instant instant.

\noindent The $b$-th ($b=1,...,B$) field reflected at $t^{\left(c\right)}$
($c=1,...,C$) by $\Xi$ ($\Xi\triangleq\Xi_{RIS}\cup\Xi_{wall}$)
in far-field is given by \cite{Oliveri 2021c}\cite{Oliveri 2022}\cite{Oliveri 2022d}\begin{equation}
\mathbf{F}_{b}^{\left(c\right)}\left(\mathbf{r}\right)=\frac{jk_{0}}{4\pi}\frac{\exp\left(-jk_{0}\left|\mathbf{r}\right|\right)}{\left|\mathbf{r}\right|}\int_{\Xi}\left\{ \mathbf{J}_{b}^{\left(c\right)}\left(\widetilde{\mathbf{r}}\right)\exp\left(jk_{0}\widehat{\mathbf{r}}\cdot\widetilde{\mathbf{r}}\right)\right\} \mathrm{d}\widetilde{\mathbf{r}}\label{eq:reflected field}\end{equation}
where $\mathbf{J}_{b}^{\left(c\right)}$ is the $b$-th ($b=1,...,B$)
surface (i.e., $\mathbf{r}\in\Xi$) equivalent current equal to\begin{equation}
\mathbf{J}_{b}^{\left(c\right)}\left(\mathbf{r}\right)=\widehat{\mathbf{r}}\times\left[\eta_{0}\widehat{\mathbf{r}}\times\mathbf{J}_{b,e}^{\left(c\right)}\left(\mathbf{r}\right)+\mathbf{J}_{b,h}^{\left(c\right)}\left(\mathbf{r}\right)\right],\label{eq:}\end{equation}
$k_{0}$ and $\eta_{0}$ being the free-space wave-number ($k_{0}\triangleq2\pi f_{0}\sqrt{\varepsilon_{0}\mu_{0}}$)
and the impedance ($\eta_{0}\triangleq\sqrt{\frac{\mu_{0}}{\varepsilon_{0}}}$),
respectively, while $\varepsilon_{0}$ is the free-space permittivity
and $\mu_{0}$ is the permeability $\mu_{0}$. Moreover, $\mathbf{J}_{b,o}^{\left(c\right)}$
($o\in\left\{ e,h\right\} $) is the electric/magnetic term of the
current $\mathbf{J}_{b}^{\left(c\right)}$ induced on the surface
$\Xi$.

\noindent Within the \emph{RIS} region $\Xi_{RIS}$, the current components
can be computed with the \emph{GSTC} approach \cite{Oliveri 2021c}\cite{Oliveri 2022}\cite{Oliveri 2022d}\cite{Ricoy 1990}-\cite{Senior 1987}
as follows\begin{equation}
\begin{array}{l}
\mathbf{J}_{b,e}^{\left(c\right)}\left(\mathbf{r}\right)=j2\pi f_{0}\varepsilon_{0}\left[\overline{\overline{K}}_{e}^{\left(c\right)}\left(\mathbf{r}\right)\cdot\widetilde{\mathbf{E}}_{b}^{\left(c\right)}\left(\mathbf{r}\right)\right]_{\tau}-\widehat{\bm{\nu}}\times\nabla_{\tau}\left[\overline{\overline{K}}_{h}^{\left(c\right)}\left(\mathbf{r}\right)\cdot\widetilde{\mathbf{H}}_{b}^{\left(c\right)}\left(\mathbf{r}\right)\right]_{\nu}\\
\mathbf{J}_{b,h}^{\left(c\right)}\left(\mathbf{r}\right)=j2\pi f_{0}\mu_{0}\left[\overline{\overline{K}}_{h}^{\left(c\right)}\left(\mathbf{r}\right)\cdot\widetilde{\mathbf{H}}_{b}^{\left(c\right)}\left(\mathbf{r}\right)\right]_{\tau}+\widehat{\bm{\nu}}\times\nabla_{\tau}\left[\overline{\overline{K}}_{e}^{\left(c\right)}\left(\mathbf{r}\right)\cdot\widetilde{\mathbf{E}}_{b}^{\left(c\right)}\left(\mathbf{r}\right)\right]_{\nu}\end{array}\label{eq:surface currents}\end{equation}
where $\left[\,.\,\right]_{\tau/\nu}$ stands for the tangential/normal
component, $\widetilde{\mathbf{E}}_{b}^{\left(c\right)}\left(\mathbf{r}\right)$
($\widetilde{\mathbf{H}}_{b}^{\left(c\right)}\left(\mathbf{r}\right)$)
is the local surface averaged electric (magnetic) field (i.e., $\widetilde{\mathbf{E}}_{b}^{\left(c\right)}\left(\mathbf{r}\right)=\mathcal{F}_{e}\left\{ \overline{\overline{K}}_{e}^{\left(c\right)}\left(\mathbf{r}\right);\,\mathbf{E}_{b}^{\left(c\right)}\left(\mathbf{r}\right)\right\} $,
$\widetilde{\mathbf{H}}_{b}^{\left(c\right)}\left(\mathbf{r}\right)=\mathcal{F}_{h}\left\{ \overline{\overline{K}}_{h}^{\left(c\right)}\left(\mathbf{r}\right);\,\mathbf{H}_{b}^{\left(c\right)}\left(\mathbf{r}\right)\right\} $
\cite{Oliveri 2021c}\cite{Oliveri 2022}\cite{Oliveri 2022d}), and
$\widehat{\bm{\nu}}$ is the outward normal to the \emph{RIS} surface
$\Xi_{\mathrm{RIS}}$. Moreover, $\overline{\overline{K}}_{e}^{\left(c\right)}$
($\overline{\overline{K}}_{h}^{\left(c\right)}$) is the electric
(magnetic) local surface susceptibility tensor that describes the
micro-scale electromagnetic response of the \emph{RIS}.

\noindent According to the \emph{GSTC} approach \cite{Oliveri 2021c}\cite{Oliveri 2022}\cite{Oliveri 2022d}\cite{Ricoy 1990}-\cite{Senior 1987}
and subject to the local periodicity condition \cite{Oliveri 2021c}\cite{Oliveri 2022d}\cite{Yang 2019},
the $o$-th ($o\in\left\{ e,h\right\} $) tensor within the \emph{RIS}
(i.e., $\mathbf{r}\in\Xi_{RIS}$) can be expressed as

\noindent \begin{equation}
\overline{\overline{K}}_{o}^{\left(c\right)}\left(\mathbf{r}\right)\triangleq\sum_{p=1}^{P}\left[\sum_{d=x,y,z}K_{o}^{d}\left(\underline{g};\, s_{p}^{\left(c\right)}\right)\widehat{\mathbf{d}}\widehat{\mathbf{d}}\right]\Omega_{p}\left(\mathbf{r}\right)\label{eq:susceptibilities}\end{equation}
where $\Omega_{p}$ is the basis function related to the $p$-th ($p=1,...,P$)
meta-atom that occupies the area $\Xi_{p}$ of the \emph{RIS} ($\Xi_{RIS}=\cup_{p=1}^{P}\left\{ \Xi_{p}\right\} $)
($\Omega_{p}\left(\mathbf{r}\right)$ $=$ \{$1$ if $\mathbf{r}\in\Xi_{p}$;
$0$ otherwise\}), $\underline{g}$ is the $U$-sized vector ($\underline{g}=\left\{ g^{\left(u\right)};\, u=1,...,U\right\} $)
of the micro-scale geometrical/material descriptors of the unit cell
of the \emph{RIS} featuring $\mathcal{B}$ reconfiguration bits, and
$s_{p}^{\left(c\right)}$ is the micro-scale \emph{}(discrete) status
of the $p$-th ($p=1,...,P$) \emph{RIS} atom at the $c$-th ($c=1,...,C$)
time-instant ($s_{p}^{\left(c\right)}\in\left\{ 1,...,2^{\mathcal{B}}\right\} $).

\noindent As for the wall region $\Xi_{wall}$, equations (\ref{eq:surface currents})
still hold true provided that the area $\Xi_{\mathrm{wall}}$ is partitioned
in $Q$ sub-regions (i.e., $\Xi_{wall}=\cup_{q=1}^{Q}\left\{ \Xi_{q}\right\} $)
to compute (\ref{eq:susceptibilities}) and the associated electric/magnetic
local surface susceptibilities are set to a constant value over time
(i.e., $\overline{\overline{K}}_{e}^{\left(c\right)}\leftarrow\overline{\overline{K}}_{e}$,
$\overline{\overline{K}}_{h}^{\left(c\right)}\leftarrow\overline{\overline{K}}_{h}$)
since the supporting wall properties cannot be re-configured.

\noindent According to this formulation (\ref{eq:reflected field})-(\ref{eq:susceptibilities}),
once the meta-atom has been designed {[}i.e., $\underline{g}$ is
set in (\ref{eq:susceptibilities}){]}, the $c$-th ($c=1,...,C$)
far-field pattern $\mathbf{F}_{b}^{\left(c\right)}$ reflected by
the surface $\Xi$ when illuminated by the $b$-th ($b=1,...,B$)
\emph{BS} beam can be adaptively controlled by acting on the $M$
\emph{BS} excitations affording the $b$-th ($b=1,...,B$) beam, $\underline{\alpha}_{b}^{\left(c\right)}=\left\{ \alpha_{m,b}^{\left(c\right)};\, m=1,...,M\right\} $
and/or the status of the $P$ meta-atoms of the \emph{RIS}, $\underline{s}^{\left(c\right)}\triangleq\left\{ s_{p}^{\left(c\right)};\, p=1,...,P\right\} $.

\noindent Under the assumptions of \cite{Oliveri 2019e}\cite{Oliveri 2022e}
(\emph{i}) single-carrier band-pass digitally-modulated signals at
the transmitters observed for the duration of a single pulse, (\emph{ii})
equal power distribution among the down-link beams, (\emph{iii}) mutual
incoherence among the signal and the noise for each beam, (\emph{iv})
ideal isotropic antenna in reception, the \emph{MIMO} down-link throughput
at the $l$-th ($l=1,...,L$) receiver in the $c$-th ($c=1,...,C$)
time-instant, $\mathcal{T}_{l}^{\left(c\right)}$, is given by \cite{Oliveri 2019e}\cite{Oliveri 2022e}\begin{equation}
\mathcal{T}_{l}^{\left(c\right)}=\log_{2}\left(1+\frac{\left|\mathbf{F}_{b}^{\left(c\right)}\left(\mathbf{r}_{l}^{\left(c\right)}\right)\right|^{2}}{\sum_{b=1,b\neq l}^{B}\left|\mathbf{F}_{b}^{\left(c\right)}\left(\mathbf{r}_{l}^{\left(c\right)}\right)\right|^{2}+\frac{L\sigma^{2}}{\Lambda}}\right)\label{eq:capacity beam}\end{equation}
where $\Lambda$ is the total power radiated by the \emph{BS} and
$\sigma^{2}$ is the noise power at $f_{0}$, which is assumed to
be constant across all the $L$ receivers for the sake of simplicity. 

\noindent For a fixed meta-atom geometry, $\underline{g}$, the maximization
of the minimum \emph{QoS} across all $L$ users corresponds to the
minimization of the \emph{QoS} cost function $\Phi$ defined as

\noindent \begin{equation}
\Phi\left(\underline{s}^{\left(c\right)},\, A^{\left(c\right)}\right)=\frac{1}{\mathcal{T}_{\mathrm{worst}}^{\left(c\right)}}\label{eq:W rsciugni}\end{equation}
where $\mathcal{T}_{\mathrm{worst}}^{\left(c\right)}\triangleq\min_{l=1,...,L}\left[\mathcal{T}_{l}^{\left(c\right)}\right]$. 

\noindent Such a cost function (\ref{eq:W rsciugni}) depends on the
$c$-th ($c=1,...,C$) set of the $B$ beamforming excitations of
the \emph{BS}, $A^{\left(c\right)}\triangleq\left\{ \underline{\alpha}_{b}^{\left(c\right)};\, b=1,...,B\right\} $,
and on the $c$-th ($c=1,...,C$) \emph{RIS} configuration, $\underline{s}^{\left(c\right)}$.
However, to follow the guidelines stated in Sect. \ref{sec:Introduction},
the \emph{ZF} beamforming method is assumed at the \emph{BS} \cite{Oliveri 2019e}\cite{Oliveri 2022e}
so that the resulting \emph{RISC} algorithm can blend into existing
wireless networks without modifying the \emph{BS} operation. More
specifically, the optimal beamforming coefficients, $A_{\mathrm{ZF}}^{\left(c\right)}$,
are obtained \cite{Oliveri 2019e}\cite{Oliveri 2022e} as the pseudo-inverse
of the discretized version of the overall {}``cascaded'' \cite{Swindlehurst 2022}
Green's matrix from the \emph{BS} to the user terminals\begin{equation}
\Upsilon^{\left(c\right)}\triangleq\left\{ \upsilon_{l,m}^{\left(c\right)};\, m=1,...,M;\, l=1,...,L\right\} ,\label{eq:}\end{equation}
whose ($l$, $m$)-th entry is given by\begin{equation}
\upsilon_{l,m}^{\left(c\right)}=\left.\mathbf{F}_{b}^{\left(c\right)}\left(\mathbf{r}_{l}^{\left(c\right)}\right)\right\rfloor _{\mathbf{E}_{b}^{\left(c\right)}\left(\mathbf{\mathbf{r}_{l}^{\left(c\right)}}\right)=\mathcal{G}\left(\mathbf{r}_{m},\mathbf{r}_{l}^{\left(c\right)}\right)\alpha_{m,b}^{\left(c\right)}\widehat{\bm{\chi}}}.\label{eq:cascaded matrix}\end{equation}
As for the computation of $A_{\mathrm{ZF}}^{\left(c\right)}$, it
only requires the measurement of $\Upsilon^{\left(c\right)}$ from
the \emph{BS}\begin{equation}
A_{\mathrm{ZF}}^{\left(c\right)}\triangleq\left[\Upsilon^{\left(c\right)}\right]^{\dagger},\label{eq:ZF weights}\end{equation}
$\left[\cdot\right]^{\dagger}$ being the Moore-Penrose pseudo-inverse
operator. It is worth pointing out that, even though $\Upsilon^{\left(c\right)}$
depends on the $c$-th ($c=1,...,C$) \emph{RIS} status, $\underline{s}^{\left(c\right)}$,
according to (\ref{eq:cascaded matrix}), no cooperation or sensing
from the \emph{RIS} is needed to implement the proposed \emph{RISC}
scheme.

\noindent Taking into account all previous considerations, the \emph{RISC}
problem at hand can be finally re-formulated into the following dynamic
optimization one

\begin{quotation}
\noindent \emph{Dynamic RISC Problem} - At each $c$-th ($c=1,...,C$)
time-instant, given $\underline{g}$ and $A_{\mathrm{ZF}}^{\left(c\right)}$,
find $\underline{s}_{opt}^{\left(c\right)}$ such that\begin{equation}
\Phi\left(\underline{s}^{\left(c\right)}\right)=\left.\Phi\left(\underline{s}^{\left(c\right)},\, A^{\left(c\right)}\right)\right\rfloor _{A^{\left(c\right)}=A_{\mathrm{ZF}}^{\left(c\right)}}\label{eq:cost function}\end{equation}
is minimized (i.e., $\underline{s}_{opt}^{\left(c\right)}$ $=$ $\arg$
$\left[\min_{\underline{s}^{\left(c\right)}}\Phi\left(\underline{s}^{\left(c\right)}\right)\right]$).
\end{quotation}
\noindent In this problem, the solution space is discrete and the
number of solutions $\mathbb{D}$ grows exponentially with both the
number of \emph{RIS} atoms, $P$, and the number of bits-per-atom,
$\mathcal{B}$ ($\mathbb{D}=2^{\mathcal{B}\times P}$).

\section{\noindent Memory-Enhanced \emph{RIS} Control Algorithm\label{sec:Control}}

\noindent In principle, the solution of the dynamic \emph{RISC} problem
(Sect. \ref{sec:Problem-Formulation}) may be carried out independently
for every $c$-th ($c=1,...,C$) time-instant and one instant at a
time, but it would neglect the exploitation of the spatial coherence
of the time-evolution of the terminal positions and of the resulting
$\Upsilon^{\left(c\right)}$ due to the finite speed of any user.
Moreover, the typical movement of both pedestrians and vehicles follow
repeated {}``patterns'' especially in urban indoor and outdoor scenarios
\cite{Grigorev 2019}-\cite{Zhang 2020}. Such observations suggest
that the \emph{RISC} methods that exploit the correlation between
different scenarios could more efficiently identify the optimal \emph{RIS}
configuration. Furthermore, the implementation of our \emph{RISC}
method must account for (\emph{i}) the non-linearity of the cost function
(\ref{eq:cost function}), (\emph{ii}) the discrete nature of the
solution descriptors, $\underline{s}^{\left(c\right)}$, and (\emph{iii})
the size of the solution space, $\mathbb{D}$. A natural choice to
address these challenges is that of recurring to memory-enhanced (\emph{ME})
evolutionary techniques based on Genetic paradigms \cite{Caorsi 2003}\cite{Massa 2004}.
As a matter of fact,

\begin{itemize}
\item \noindent such class of methods has been successfully adopted for
the control of complex \emph{EM} devices operating in dynamic environments
such as phased arrays \cite{Caorsi 2003}\cite{Massa 2004};
\item the multi-minima and discrete nature of the optimization problem at
hand is known to be effectively tackled by Genetic-inspired methodologies
\cite{Rocca 2009};
\item the underlying paradigm can be customized to account for arbitrary
additional constraints and goals without requiring disruptive methodological
modifications \cite{Caorsi 2003}\cite{Massa 2004}.
\end{itemize}
\noindent Accordingly, a memory-enhanced dynamic evolutionary \emph{RISC}
method (\emph{ME-RISC}) is proposed in the following according to
the flowchart in Fig. 2. Towards this end, the algorithm proposed
in \cite{Caorsi 2003}\cite{Massa 2004} is extended here to the \emph{Dynamic
RISC} problem at hand.

\noindent Following the Genetic Algorithm (\emph{GA}) guidelines \cite{Rocca 2009},
a multi-step process is performed at each $c$-th ($c=1,...,C$) instant,
$t^{\left(c\right)}$, by iteratively applying the Genetic operators
to the population of $\Gamma$ trial solutions $\mathcal{S}_{v}^{\left(c\right)}$
($\mathcal{S}_{v}^{\left(c\right)}=\left\{ \left.\underline{s}^{\left(c\right)}\right|_{v}^{\gamma};\gamma=1,...,\Gamma\right\} $,
$\gamma$ and $v$ being the solution index and the iteration index,
respectively). At the beginning of the $c$-th ($c=1,...,C$) optimization
step, an initial population $\mathcal{S}_{0}^{\left(c\right)}$ is
randomly-generated by setting its $\gamma$-th ($\gamma=1,...,\Gamma$)
individual as follows\begin{equation}
\left.\underline{s}^{\left(c\right)}\right|_{0}^{\gamma}\triangleq\left\{ \left.s_{p}^{\left(c\right)}\right|_{0}^{\gamma}=\mathrm{rnd}\left[1,2^{\mathcal{B}}\right];\, p=1,...,P\right\} ,\label{eq:}\end{equation}
then the \emph{ME-GA} loop starts ($v=1$). Firstly, the \emph{Mutation}
and the \emph{Crossover} operators are applied to yield the $v$-th
population $\mathcal{S}_{v}^{\left(c\right)}$ \cite{Massa 2004}.
Unlike standard \emph{GA}s and to increase the population diversity,
the mutation probability $\rho_{v}^{\left(c\right)}$ and the crossover
probability $\psi_{v}^{\left(c\right)}$ are dynamically updated in
the range $\rho_{v}^{\left(c\right)}\in\left[\rho_{\min},\rho_{\max}\right]$
and $\psi_{v}^{\left(c\right)}\in\left[\psi_{\min},\psi_{\max}\right]$,
respectively, by means of the following update rules \cite{Massa 2004}\begin{equation}
\begin{array}{c}
\rho_{v}^{\left(c\right)}=\rho_{\max}-\left(\rho_{\max}-\rho_{\min}\right)\frac{\sigma_{v}^{\left(c\right)}}{\sigma_{\max}}\\
\psi_{v}^{\left(c\right)}=\psi_{\min}+\left(\psi_{\max}-\psi_{\min}\right)\frac{\sigma_{v}^{\left(c\right)}}{\sigma_{\max}}\end{array}\label{eq:}\end{equation}
where $\sigma_{v}^{\left(c\right)}$ is the variance of the \emph{}population
at the $v$-th iteration\begin{equation}
\sigma_{v}^{\left(c\right)}=\frac{1}{\Gamma}\sum_{\gamma=1}^{\Gamma}\left|\left.\underline{s}^{\left(c\right)}\right|_{v}^{\gamma}-\underline{s}_{avg\left(v\right)}^{\left(c\right)}\right|^{2},\label{eq:}\end{equation}
$\sigma_{\max}$ being its maximum, while $\underline{s}_{avg\left(v\right)}^{\left(c\right)}=\frac{1}{\Gamma}\sum_{\gamma=1}^{\Gamma}\left.\underline{s}^{\left(c\right)}\right|_{v}^{\gamma}$.

\noindent The fitness of all the individuals of the current $v$-th
population is then evaluated by computing, for each $\gamma$-th ($\gamma=1,...,\Gamma$)
one, the corresponding cost function value (\ref{eq:cost function})
(i.e., $\Phi_{v}^{\gamma}\triangleq\Phi\left(\left.\underline{s}^{\left(c\right)}\right|_{v}^{\gamma}\right)$)
and the $v$-th iteration best individual, $\underline{s}_{best\left(v\right)}^{\left(c\right)}$,
along with the associated fitness value, $\Phi_{best\left(v\right)}^{\left(c\right)}$,
are deduced {[}i.e., $\underline{s}_{best\left(v\right)}^{\left(c\right)}=\arg\left[\min_{\gamma=1,..,\Gamma;\widetilde{v}=1,...,v}\left\{ \Phi\left(\left.\underline{s}^{\left(c\right)}\right|_{\widetilde{v}}^{\gamma}\right)\right\} \right]$
and $\Phi_{best\left(v\right)}^{\left(c\right)}\triangleq\Phi\left(\underline{s}_{best\left(v\right)}^{\left(c\right)}\right)${]}.

\noindent To leverage on the memory acquired by the \emph{EA} during
operation, a set of non-conventional Genetic operators is then applied.
More in detail, firstly a fraction $\nu_{v}^{\left(c\right)}$ {[}$\nu_{v}^{\left(c\right)}=\nu_{\max}\left(1-\frac{\sigma_{v}^{\left(c\right)}}{\sigma_{\max}}\right)${]}
of the individuals of the population $\mathcal{S}_{v}^{\left(c\right)}$
with worse fitnesses are replaced with randomly sampled individuals
\cite{Massa 2004} ({}``\emph{Population Replacement}'' step - Fig.
2). Successively, the indicator $\Theta_{v}^{\left(c\right)}$ is
computed\begin{equation}
\Theta_{v}^{\left(c\right)}=\sum_{\widetilde{v}=1}^{\mathcal{W}}\frac{\Phi_{best\left(v\right)}^{\left(c\right)}-\Phi_{best\left(v-\widetilde{v}\right)}^{\left(c\right)}}{2^{\widetilde{v}}}\label{eq:W Masciulla}\end{equation}
({}``\emph{Effectiveness Self-Evaluation}'' step - Fig. 2) to quantify
the effectiveness of the optimization at the $c$-th time-instant,
$\mathcal{W}$ being a user-defined observation interval. In (\ref{eq:W Masciulla})
and unlike \cite{Massa 2004}, the exponentially-decaying weighting
$2^{\widetilde{v}}$ is considered to improve the responsiveness of
the \emph{RISC} method to changes in the dynamic scenario.

\noindent Depending on the value of $\Theta_{v}^{\left(c\right)}$,
either the {}``\emph{Memory Learning}'' step ($\Theta_{v}^{\left(c\right)}\geq0$)
or the {}``\emph{Memory Remembering}'' step ($\Theta_{v}^{\left(c\right)}<0$)
is performed. In the former case, the best individual at the current
$v$-th iteration, $\underline{s}_{best\left(v\right)}^{\left(c\right)}$,
is stored into a memory pool of good solutions, $\Psi$ (denoted hereinafter
as the {}``\emph{memory}''), with probability $\beta_{v}^{\left(c\right)}$
given by\begin{equation}
\beta_{v}^{\left(c\right)}=\frac{\Theta_{v}^{\left(c\right)}}{\Theta_{max}^{\left(c\right)}}\beta_{\max},\label{eq:}\end{equation}
$\beta_{\max}$ being the maximum probability of storing and $\Theta_{\max}^{\left(c\right)}=\max_{\widetilde{v}=1,..,v}\left[\Theta_{\widetilde{v}}^{\left(c\right)}\right]$.
In the second case, the worst individual of the current population
$\mathcal{S}_{v}^{\left(c\right)}$, $\underline{s}_{worst\left(v\right)}^{\left(c\right)}$
($\underline{s}_{worst\left(v\right)}^{\left(c\right)}$ $\triangleq$
$\arg$ {[} $\max$$_{\gamma=1,..,\Gamma;\widetilde{v}=1,...,v}$
\{$\Phi\left(\left.\underline{s}^{\left(c\right)}\right|_{\widetilde{v}}^{\gamma}\right)$\}{]}),
is substituted by the best individual of the \emph{memory} $\Psi$
with a probability $\kappa_{v}^{\left(c\right)}$ equal to\begin{equation}
\kappa_{v}^{\left(c\right)}=\frac{\Theta_{v}^{\left(c\right)}}{\Theta_{max}^{\left(c\right)}}\kappa_{\max},\label{eq:}\end{equation}
$\kappa_{\max}$ being the maximum probability of re-storing.

\noindent If $v=V$ or a stagnation condition over the sliding window
of length $\mathcal{W}$ is detected (i.e., $\left|\Phi_{best\left(v\right)}^{\left(c\right)}-\frac{1}{\mathcal{W}}\sum_{\widetilde{v}=1}^{\mathcal{W}}\Phi_{best\left(v-\widetilde{v}\right)}^{\left(c\right)}\right|<\delta$,
$\delta$ being the stagnation threshold), then the optimization process
is stopped and the $c$-th \emph{RIS} setup is outputted by setting
$\underline{s}_{opt}^{\left(c\right)}=\underline{s}_{best\left(v\right)}^{\left(c\right)}$.
Otherwise, the iteration index is updated ($v\leftarrow v+1$) and
the \emph{ME-GA} loop is repeated starting from the application of
the \emph{Mutation} and \emph{Crossover} operators.

\section{\noindent Numerical Results\label{sec:Results}}

\noindent To illustrate the performance and to assess the effectiveness/reliability
of the proposed \emph{ME-RISC} method, selected numerical results
are reported and discussed in the following with comparisons, as well.

\noindent The benchmark scenario refers to a \emph{BS} operating at
$f_{0}=3.5$ {[}GHz{]} and composed by $M=30\times30$ half-wavelength
spaced single-polarization antennas. The \emph{BS} is located $5$
{[}m{]} above the ground and it illuminates a \emph{RIS} centered
at $5$ {[}m{]} from the ground in a concrete wall of area $\Xi_{wall}=6\times7$
{[}$\textnormal{m}^{2}${]} and thickness $0.2$ {[}m{]}. The \emph{RIS}
features a side of $\ell_{RIS}\approx1.93$ {[}m{]} and it consists
of $P=45\times45$ reconfigurable meta-atoms ($P=2025$) whose micro-scale
geometrical/material descriptors (i.e., $\underline{g}$) have been
designed by scaling to $f_{0}$ the $\mathcal{B}$-bit ($\mathcal{B}=3$)
unit cell described in \cite{Liang 2022}. The wall has been modeled
by setting its relative permittivity to $\varepsilon_{wall}=5.24$
and the conductivity to $0.123$ {[}S/m{]} \cite{ITU 2021}. Moreover,
according to the guidelines in \cite{Zhang 2021a}, the \emph{BS}
power and the environment noise level have been set to $\Lambda=46$
{[}dBm{]} and $\sigma^{2}=-96$ {[}dBm{]}, respectively. Finally,
the setup of the control parameters of the \emph{ME-RISC} has been
chosen according to the literature guidelines in \cite{Massa 2004}\cite{Rocca 2009}:
$\sigma_{\max}=1$, $\left[\rho_{\min},\rho_{\max}\right]=\left[0.02,0.06\right]$,
$\left[\psi_{\min},\psi_{\max}\right]=\left[0.6,0.95\right]$, $\mathcal{W}=3$,
$\nu_{\max}=\kappa_{\max}=\beta_{\max}=0.2$, $V=100$, and $\Gamma=100$.

\noindent The first numerical experiment is aimed at illustrating
the working of the \emph{ME-RISC} method. Towards this end, the dynamic
scenario in Fig. 3 is dealt with where a \emph{BS}, placed at the
coordinates $\left(x',y',z'\right)=\left(-20,20,5\right)$ {[}m{]}
of the global reference system, serves $L=3$ users moving for $C=100$
subsequent time-instants according to an {}``Aperiodic Trajectory''
in an area of $\Omega_{user}=120\times80$ {[}$\textnormal{m}^{2}${]}
in front of the \emph{RIS}. For illustrative purposes, Figure 4(\emph{a})
gives an idea of the standard behaviour of the \emph{ME-GA} operators
during the iterative process at the basis of the \emph{ME-RISC} method.
More specifically, the plots of the self-detected efficiency indicator
$\Theta_{v}^{\left(c\right)}$, the store probability $\beta_{v}^{\left(c\right)}$,
and the restore probability $\kappa_{v}^{\left(c\right)}$ during
a period of $3$ successive time instants are shown. As it can be
observed, the sign/value of $\Theta_{v}^{\left(c\right)}$ forces
either the exploitation of the memory of the \emph{ME-RISC} {[}i.e.,
the {}``Individual Restored'' event - Fig. 4(\emph{a}){]} or the
storage of the current best solution into the memory $\Psi${[}i.e.,
the {}``Individual Stored'' event - Fig. 4(\emph{a}){]}. As a result
of the memory use, the worst \emph{MIMO} down-link throughput, $\mathcal{T}_{worst}^{\left(c\right)}$,
yielded by the \emph{ME-RISC} turns \emph{}out to be always (i.e.,
$\forall c\in\left[1,\, C\right]$) greater than that when applying
a standard \emph{GA} without memory-based operators ({}``\emph{GA}-\emph{RISC}'')
(i.e., $\left.\mathcal{T}_{worst}^{\left(c\right)}\right|_{ME-RISC}>\left.\mathcal{T}_{worst}^{\left(c\right)}\right|_{GA-RISC}$
- Fig. 4(\emph{b}){]}.

\noindent The motivation for such a performance improvement can be
inferred by the plot of the difference between the optimal configuration
of the \emph{RIS} at a $c$-th ($c=1,...,C$) time-instant, $\underline{s}_{opt}^{\left(c\right)}$,
and that at the initial iteration (i.e., $v=1$) of the \emph{ME-GA}
algorithm in the same instant, $\underline{s}_{best\left(1\right)}^{\left(c\right)}$\begin{equation}
\Delta\underline{s}^{\left(c\right)}=\left|\underline{s}_{best\left(1\right)}^{\left(c\right)}-\underline{s}_{opt}^{\left(c\right)}\right|.\label{eq:}\end{equation}
By comparing the color-maps of $\Delta\underline{s}_{ME-RISC}^{\left(c\right)}$
{[}Fig. 5(\emph{a}){]} and $\Delta\underline{s}_{GA-RISC}^{\left(c\right)}$
{[}Fig. 5(\emph{b}){]} at a representative time-instant (e.g., $c=20)$,
one can deduce that the {}``memory'' mechanism allows an initialization
of the \emph{RIS} configuration closer to the optimal one (i.e., $\Delta s_{p}^{\left(c\right)}\to0$,
$p=1,...,P$), hence reducing the optimization burden/time and speeding-up
the convergence to the optimal \emph{RIS} setup.

\noindent The advantage of deploying a \emph{RIS} controlled by the
\emph{ME-RISC} is further highlighted by the plots of the \emph{MIMO}
down-link throughput at the $L$ receiver during the $C$ time-steps,
$\mathcal{T}_{l}^{\left(c\right)}$ ($c=1,...,C$), $l=1,...,L$,
shown in Fig. 6. Whatever the $l$-th ($l=1,...,L$) the user, the
throughput is improved (i.e., $\left.\mathcal{T}_{l}^{\left(c\right)}\right|_{w/\, RIS}>\left.\mathcal{T}_{l}^{\left(c\right)}\right|_{w/o\, RIS}$,
$l=1,...,L$) and the control strategy yields a \emph{QoS} fairness
among all users (i.e., $\left.\mathcal{T}_{1}^{\left(c\right)}\right|_{w/\, RIS}\approx\left.\mathcal{T}_{l}^{\left(c\right)}\right|_{w/\, RIS}\approx\left.\mathcal{T}_{L}^{\left(c\right)}\right|_{w/\, RIS}$
). 

\noindent For completeness, Figure 7 gives the power footprint $\mathcal{P}$
(i.e., $\mathcal{P}_{b}^{\left(c\right)}\left(x',y'\right)\triangleq\left|\mathbf{F}_{b}^{\left(c\right)}\left(\mathbf{r}\right)\right|_{z'=0}^{2}$
, $\mathbf{r}\in\Omega_{user}$) at the $c$-th ($c=20$) instant
when the \emph{BS} operates in the scenario without the \emph{RIS}
{[}Fig. 7(\emph{a}), Fig. 7(\emph{c}), and Fig. 7(\emph{e}){]} or
with the \emph{RIS} {[}Fig. 7(\emph{b}), Fig. 7(\emph{d}), and Fig.
7(\emph{f}){]} and it serves the user $l=1$ {[}Figs. 7(\emph{a})-7(\emph{b}){]},
the user $l=2$ {[}Figs. 7(\emph{c})-7(\emph{d}){]}, and the user
$l=L=3$ {[}Figs. 7(\emph{e})-7(\emph{f}){]}. As expected, adding
the \emph{ME-RISC}-controlled \emph{RIS} better focus the power transmitted
by the \emph{BS} towards the user to serve {[}e.g., $\left.\mathcal{P}_{l}^{\left(c\right)}\left(\mathbf{r}_{l}^{\left(c\right)}\right)\right|_{w/\, RIS}>\left.\mathcal{P}_{l}^{\left(c\right)}\left(\mathbf{r}_{l}^{\left(c\right)}\right)\right|_{w/o\, RIS}$
- ($l=1$) Figs. 7(\emph{a})-7(\emph{b}); ($l=2$) Figs. 7(\emph{c})-7(\emph{d});
and ($l=L=3$) - Figs. 7(\emph{e})-7(\emph{f}){]}.

\noindent Previous outcomes on the benefit of enriching the propagation
environment with a \emph{ME-RISC}-controlled \emph{RIS} are confirmed
also when dealing with different noise levels. The time evolution
of $\mathcal{T}_{worst}^{\left(c\right)}$ ($c=1,...,C$; $C=100)$
for different noise levels (i.e., $\sigma^{2}\in\left\{ -96,-76,-56\right\} $
{[}dBm{]}) in Fig. 8(\emph{a}) shows that there is always a considerable
improvement of the worst-case throughput {[}e.g., $\left.\mathcal{T}_{worst}^{\left(c\right)}\right|_{\mathrm{w/o\, RIS}}^{\sigma^{2}}<\left.\mathcal{T}_{worst}^{\left(c\right)}\right|_{\mathrm{w/\, RIS}}^{\sigma^{2}}$
and $\left.\mathcal{T}_{worst}^{\left(c\right)}\right|_{\mathrm{w/o\, RIS}}^{\sigma^{2}=-96\,\mathrm{dBm}}\approx\left.\mathcal{T}_{worst}^{\left(c\right)}\right|_{\mathrm{w/\, RIS}}^{\sigma^{2}=-76\,\mathrm{dBm}}$
- Fig. 8(\emph{a}){]} as visually further pointed out by the behavior
of the average worst-case throughput $\mathcal{T}_{worst}^{avg}$
($\mathcal{T}_{worst}^{avg}\triangleq\frac{1}{C}\sum_{c=1}^{C}\mathcal{T}_{worst}^{\left(c\right)}$)
versus the noise level {[}Fig. 8(\emph{b}){]}. This latter proves
the boost of the \emph{MIMO} communication performance in all operative
conditions {[}$\left.\mathcal{T}_{worst}^{avg}\right|_{w/\, RIS}>\left.\mathcal{T}_{worst}^{avg}\right|_{w/o\, RIS}$
- Fig. 8(\emph{b}){]}\@.

\noindent Besides the contribution of the \emph{RIS} installation,
it is worth remarking that the proposed \emph{ME-RISC} also opportunistically
leverages on the scattering environment - according to the \emph{SEME}
paradigm - to enhance the \emph{QoS}. To assess such a feature, another
(even more challenging) scenario is considered where both $\Xi_{RIS}$
and $\Xi_{wall}$ have been shifted by $10$ {[}m{]} along the $x'$
axis with respect to the setup of the first numerical experiment.
Such a choice implies that the specular reflection from the wall is
towards one of the users (i.e., $l=2$) making more difficult the
task of the \emph{RIS} to provide a good customer service (i.e., an
acceptable throughput to all $L$ users).

\noindent The plots of $\mathcal{T}_{\mathrm{worst}}^{\left(c\right)}$
($1\le c\le C$; $C=100$) for the cases {}``\emph{w/ RIS}'', {}``\emph{w/o
RIS}'', and {}``\emph{RIS-Only}'' are shown in Fig. 9(\emph{a})
to give some useful insights to the readers. While it is confirmed
the role of the presence of the \emph{RIS} to improve the \emph{QoS}
at the users (i.e., $\left.\mathcal{T}_{worst}^{\left(c\right)}\right|_{\mathrm{w/o\, RIS}}<\left.\mathcal{T}_{worst}^{\left(c\right)}\right|_{\mathrm{w/\, RIS}}$,
$\forall\, c\in\left[1,\, C\right]$), these plots assess the effective
exploitation of the environment, since $\left.\mathcal{T}_{worst}^{\left(c\right)}\right|_{w/\, RIS}>\left.\mathcal{T}_{worst}^{\left(c\right)}\right|_{RIS-Only}$
($\forall\, c\in\left[1,\, C\right]$), which is also pictorially
highlighted by the comparison between the power footprints ($c=20$)
in Figs. 9(\emph{c})-9(\emph{d}).

\noindent The last test case does not consider aperiodic users' trajectories
(Fig. 3), but it deals with users moving according to periodic patterns
\cite{Cheshmehzangi 2012} as it generally happens to pedestrians
in a urban environment. Accordingly, the users' trajectories {[}Fig.
10(\emph{b}){]} collected in the {}``Old Market Square'' of Nottingham
(UK) \cite{Cheshmehzangi 2012} {[}Fig. 10(\emph{a}){]} have been
chosen as a realistic benchmark.

\noindent The comparison of the evolution of $\mathcal{T}_{\mathrm{worst}}^{\left(c\right)}$
in the temporal slot $90\le c\le C$ ($C=100$) when configuring the
\emph{RIS} with the {}``\emph{ME-RISC}'' method or the {}``\emph{GA}-\emph{RISC''}
one indicates once again the better performance of the proposed approach
in terms of effectiveness to yield a greater throughput as well as
convergence speed of the \emph{ME-GA} optimization loop thanks to
a suitable use of the \emph{memory} mechanism {[}i.e., $\left.\mathcal{T}_{worst}^{\left(c\right)}\right|_{ME-RISC}>\left.\mathcal{T}_{worst}^{\left(c\right)}\right|_{GA-RISC}$
- Fig. 10(\emph{a}){]}. Indeed, the repetitiveness of the communication
scenario, owing to the periodic coincidence of the locations of the
users' terminals, implies that there is an high probability that,
at each $c$-th ($c=1,...,C$) variation of the propagation conditions,
it exists a $\Psi$-stored \emph{RIS} configuration which is {}``almost
optimal'' as confirmed by the representative plot of $\Delta\underline{s}_{ME-RISC}^{\left(c\right)}$
($c=99$) in Fig. 10(\emph{b}) {[}vs. $\Delta\underline{s}_{GA-RISC}^{\left(c\right)}$
- Fig. 10(\emph{c}){]}.

\noindent For completeness, the complete ($1\le c\le C$) time-evolution
of the worst-case throughput $\mathcal{T}_{worst}^{\left(c\right)}$
with and without the \emph{RIS} is reported in Fig. 12.

\section{\noindent Conclusions and Final Remarks\label{sec:Conclusions-and-Remarks}}

\noindent A new \emph{RISC} method, called \emph{ME-RISC}, has been
proposed to dynamically configure \emph{RIS}s in \emph{MIMO} communication
scenarios by maximizing the worst-case throughput across all users.
Without requiring either a knowledge on the links from the \emph{BS}
to the \emph{RIS} and from the \emph{RIS} to the user terminals or
a modification of the beamforming strategy at the \emph{BS}, the method
leverages on the effectiveness and the efficiency of a \emph{ME} evolutionary
algorithm to optimize the states of the \emph{RIS} meta-atoms, which
are characterized by a finite number of bits and electromagnetically
modeled with the \emph{GSTC} approach.

\noindent The numerical validation has proved that (\emph{i}) the
\emph{ME-GA} algorithm, at the core of the \emph{ME-RISC} method,
outperforms standard \emph{GA}-based optimizations in terms of both
achievable \emph{QoS} and convergence speed; (\emph{ii}) the \emph{ME-RISC}
constructively exploits the \emph{RIS} as well as the scattering environment
to improve the throughput at the users' terminals by fulfilling the
key principle of the \emph{SEME} (i.e., the opportunistic exploitation
of the environment); (\emph{iii}) the \emph{ME-RISC} turns out to
be reliable and efficient in dealing with aperiodic as well as repetitive
users' behaviors, this latter framework being intrinsically the optimal
working condition owing to the memory mechanism.

\noindent Future works, beyond the scope of this manuscript, will
be aimed at validating the proposed control method when integrated
in a urban-scale \emph{EM} scenario where more {}``smart entities''
(i.e., \emph{IAB} nodes, \emph{SR}s, static passive \emph{EMS}s {[}\emph{SP-EMS}s{]},
etc ...) \cite{Yang 2022}-\cite{Oliveri 2023} coexist and they need
to be suitably planned/deployed for yielding an optimal performance/cost/maintenance
trade-off.

\section*{\noindent Acknowledgements}

\noindent This work benefited from the networking activities carried
out within the Project \char`\"{}SPEED\char`\"{} (Grant No. 6721001)
funded by National Science Foundation of China under the Chang-Jiang
Visiting Professorship Program, the Project \char`\"{}ICSC National
Centre for HPC, Big Data and Quantum Computing (CN HPC)\char`\"{}
funded by the European Union - NextGenerationEU within the PNRR Program
(CUP: E63C22000970007), the project DICAM-EXC (Departments of Excellence
2023-2027, grant L232/2016) funded by the Italian Ministry of Education,
Universities and Research (MUR), the Project \char`\"{}Smart ElectroMagnetic
Environment in TrentiNo - SEME@TN\char`\"{} funded by the Autonomous
Province of Trento (CUP: C63C22000720003), and the Project \char`\"{}AURORA
- Smart Materials for Ubiquitous Energy Harvesting, Storage, and Delivery
in Next Generation Sustainable Environments\char`\"{} funded by the
Italian Ministry for Universities and Research within the PRIN-PNRR
2022 Program. A. Massa wishes to thank E. Vico for her never-ending
inspiration, support, guidance, and help.

\newpage
\section*{FIGURE CAPTIONS}

\begin{itemize}
\item \textbf{Figure 1.} \emph{Problem geometry} - \emph{RIS}-enhanced \emph{MIMO}
down-link communication scenario.
\item \textbf{Figure 2.} \emph{ME-RISC} \emph{Method} (\emph{ME-GA Algorithm})
- Flowchart.
\item \textbf{Figure 3}. \emph{Illustrative Example} (Aperiodic Users' Trajectories;
$\Lambda=46$ {[}dBm{]}, $\sigma^{2}=-96$ {[}dBm{]}) - Scenario and
users' trajectories ($c=1,...,C$; $C=100$).
\item \textbf{Figure 4}. \emph{Illustrative Example} (Aperiodic Users' Trajectories;
$\Lambda=46$ {[}dBm{]}, $\sigma^{2}=-96$ {[}dBm{]}) - Time-evolution
of (\emph{a}) the parameters of the {}``\emph{memory}'' mechanism
and of (\emph{b}) the worst \emph{MIMO} down-link throughput $\mathcal{T}_{worst}^{\left(c\right)}$.
\item \textbf{Figure 5}. \emph{Illustrative Example} (Aperiodic Users' Trajectories;
$\Lambda=46$ {[}dBm{]}, $\sigma^{2}=-96$ {[}dBm{]}, $c=20$) - Plot
of $\Delta\underline{s}^{\left(c\right)}$ when using (\emph{a}) the
\emph{ME-RISC} and (\emph{b}) the \emph{GA-RISC} for controlling the
\emph{RIS}.
\item \textbf{Figure 6}. \emph{Illustrative Example} (Aperiodic Users' Trajectories;
$\Lambda=46$ {[}dBm{]}, $\sigma^{2}=-96$ {[}dBm{]}) - Time-evolution
of the the \emph{MIMO} down-link throughput at the $l$-th ($l=1,...,L$)
receiver, $\mathcal{T}_{l}^{\left(c\right)}$ ($c=1,...,C$; $C=100$)
with and without the \emph{RIS} controlled by the \emph{ME-RISC} method.
\item \textbf{Figure 7}. \emph{Illustrative Example} (Aperiodic Users' Trajectories;
$\Lambda=46$ {[}dBm{]}, $\sigma^{2}=-96$ {[}dBm{]}, $c=20$) - Plot
of the footprint pattern $\mathcal{P}_{b}^{\left(c\right)}\left(x',y'\right)$
when the \emph{BS} operates in the scenario (\emph{a})(\emph{c})(\emph{e})
without the \emph{RIS} or (\emph{b})(\emph{d})(\emph{f}) with the
\emph{RIS} controlled by the \emph{ME-RISC} method for serving the
user (\emph{a})(\emph{b}) $l=1$, (\emph{c})(\emph{d}) $l=2$, and
(\emph{e})(\emph{f}) $l=L=3$ ($b=l$).
\item \textbf{Figure 8}. \emph{Performance Evaluation} (Aperiodic Users'
Trajectories; $\Lambda=46$ {[}dBm{]}) - Plots of (\emph{a}) the time-evolution
($c=1,...,C$; $C=100$) of the worst \emph{MIMO} down-link throughput,
$\mathcal{T}_{worst}^{\left(c\right)}$, in correspondence with different
noise levels ($\sigma^{2}\in\left\{ -96,-76,-56\right\} $ {[}dBm{]})
and (\emph{b}) the behavior of the average worst-case throughput,
$\mathcal{T}_{worst}^{avg}$ versus the noise level $\sigma^{2}$.
\item \textbf{Figure 9}. \emph{Performance Evaluation} (Aperiodic Users'
Trajectories; $\Lambda=46$ {[}dBm{]}, $\sigma^{2}=-96$ {[}dBm{]})
- Plots of (\emph{a}) the time-evolution ($c=1,...,C$; $C=100$)
of the worst \emph{MIMO} down-link throughput, $\mathcal{T}_{worst}^{\left(c\right)}$,
and (\emph{b}) the footprint pattern $\mathcal{P}_{b}^{\left(c\right)}\left(x',y'\right)$
when the \emph{BS} operates in the scenario (\emph{b}) without the
\emph{RIS} or with the \emph{RIS} controlled by the \emph{ME-RISC}
method (\emph{c}) without or (\emph{d}) taking into account the environment
for serving the user (\emph{b})-(\emph{d}) $l=1$ ($b=l$).
\item \textbf{Figure 10}. \emph{Performance Evaluation} (Periodic Users'
Trajectories) - \emph{Old Market Square} in Nottingham (UK): (\emph{a})
photo and (\emph{b}) plot of trajectories of pedestrian users \cite{Cheshmehzangi 2012}.
\item \textbf{Figure 11}. \emph{Performance Evaluation} (Periodic Users'
Trajectories; $\Lambda=46$ {[}dBm{]}, $\sigma^{2}=-96$ {[}dBm{]})
- Plots of (\emph{a}) the time-evolution ($90\le c\le100$) of the
worst \emph{MIMO} down-link throughput $\mathcal{T}_{worst}^{\left(c\right)}$
and (\emph{b})(\emph{c}) $\Delta\underline{s}^{\left(c\right)}$ when
using (\emph{b}) the \emph{ME-RISC} or (\emph{c}) the \emph{GA-RISC}
for controlling the \emph{RIS} at the $c$-th ($c=99$) time instant.
\item \textbf{Figure 12}. \emph{Performance Evaluation} (Periodic Users'
Trajectories; $\Lambda=46$ {[}dBm{]}, $\sigma^{2}=-96$ {[}dBm{]})
- Time-evolution ($c=1,...,C$; $C=100$) of the worst \emph{MIMO}
down-link throughput, $\mathcal{T}_{worst}^{\left(c\right)}$.
\end{itemize}
~

\newpage
\begin{center}~\vfill\end{center}

\begin{center}\includegraphics[%
  width=0.95\columnwidth]{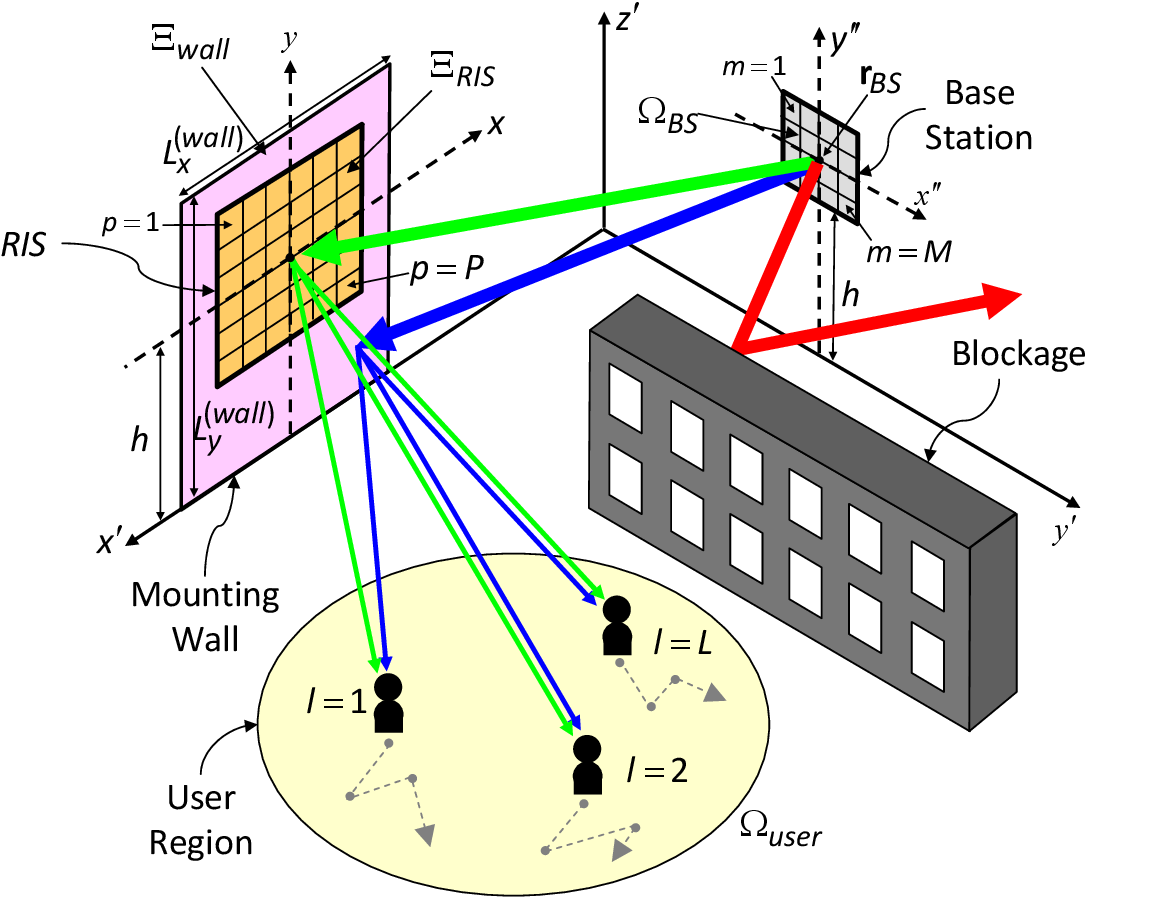}\end{center}

\begin{center}~\vfill\end{center}

\begin{center}\textbf{Fig. 1 - F. Zardi et} \textbf{\emph{al.}}\textbf{,}
\textbf{\emph{{}``}}Memory-Enhanced Dynamic Evolutionary Control...
''\end{center}

\newpage
\begin{center}~\vfill\end{center}

\begin{center}\includegraphics[%
  width=0.95\columnwidth]{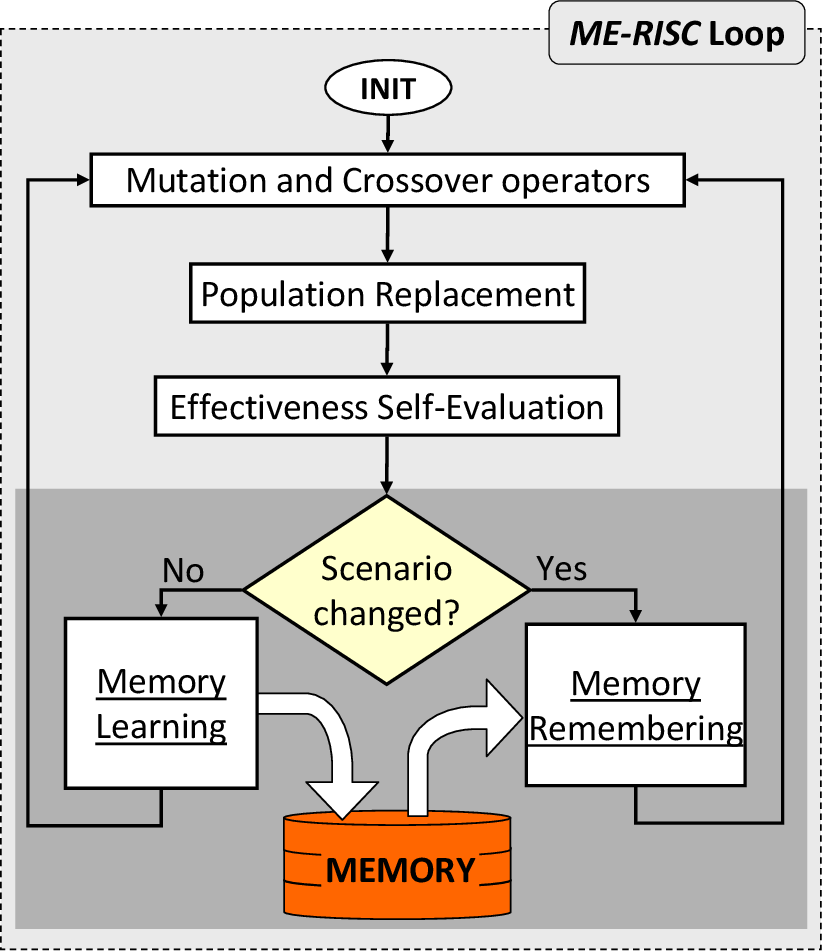}\end{center}

\begin{center}~\vfill\end{center}

\begin{center}\textbf{Fig. 2 - F. Zardi et} \textbf{\emph{al.}}\textbf{,}
\textbf{\emph{{}``}}Memory-Enhanced Dynamic Evolutionary Control...
''\end{center}

\newpage
\begin{center}~\vfill\end{center}

\begin{center}\includegraphics[%
  width=0.90\textwidth]{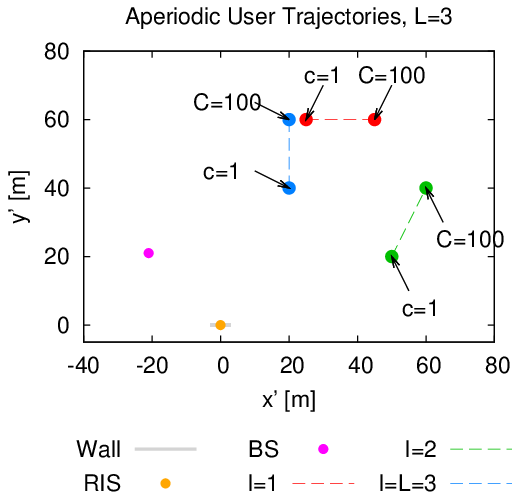}\end{center}

\begin{center}~\vfill\end{center}

\begin{center}\textbf{Fig. 3 - F. Zardi et} \textbf{\emph{al.}}\textbf{,}
\textbf{\emph{{}``}}Memory-Enhanced Dynamic Evolutionary Control...
''\end{center}

\newpage
\begin{center}~\vfill\end{center}

\begin{center}\begin{tabular}{c}
\includegraphics[%
  width=0.80\textwidth]{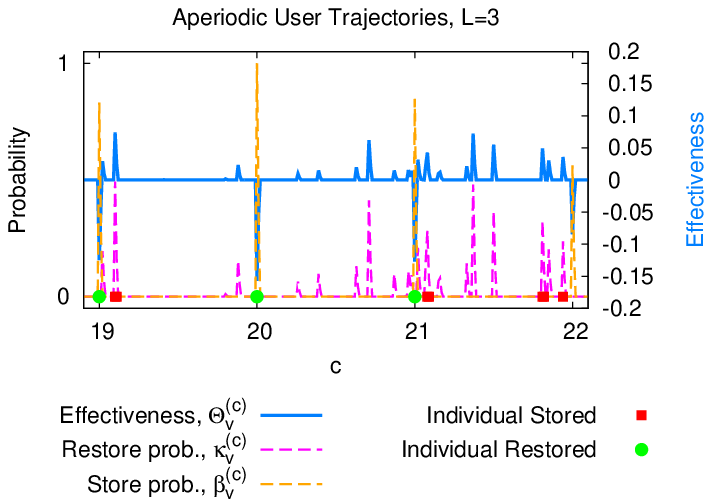}\tabularnewline
(\emph{a})\tabularnewline
\includegraphics[%
  width=0.80\textwidth]{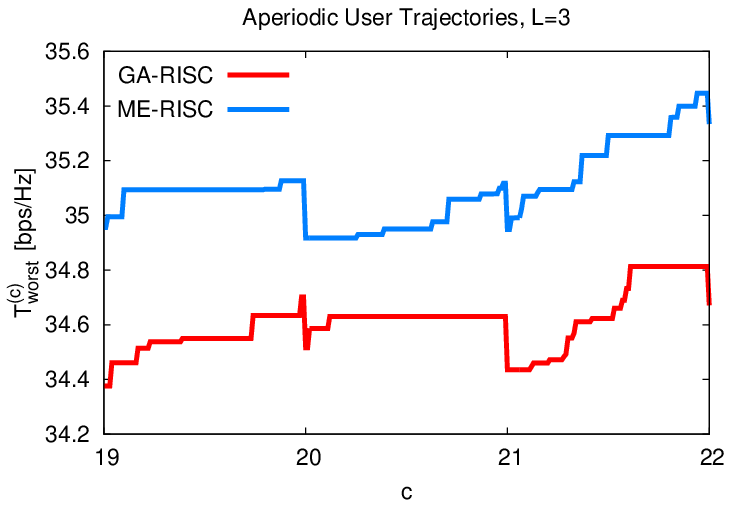}\tabularnewline
(\emph{b})\tabularnewline
\end{tabular}\end{center}

\begin{center}~\vfill\end{center}

\begin{center}\textbf{Fig. 4 - F. Zardi et} \textbf{\emph{al.}}\textbf{,}
\textbf{\emph{{}``}}Memory-Enhanced Dynamic Evolutionary Control...
''\end{center}

\newpage
\begin{center}~\vfill\end{center}

\begin{center}\begin{tabular}{c}
\includegraphics[%
  width=0.70\columnwidth]{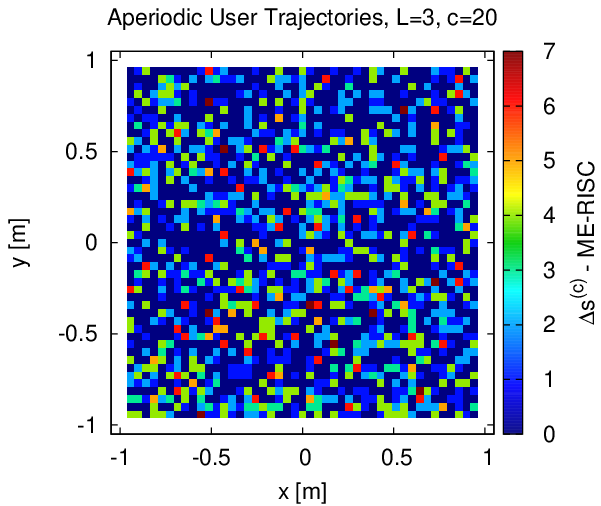}\tabularnewline
(\emph{a})\tabularnewline
\includegraphics[%
  width=0.70\columnwidth]{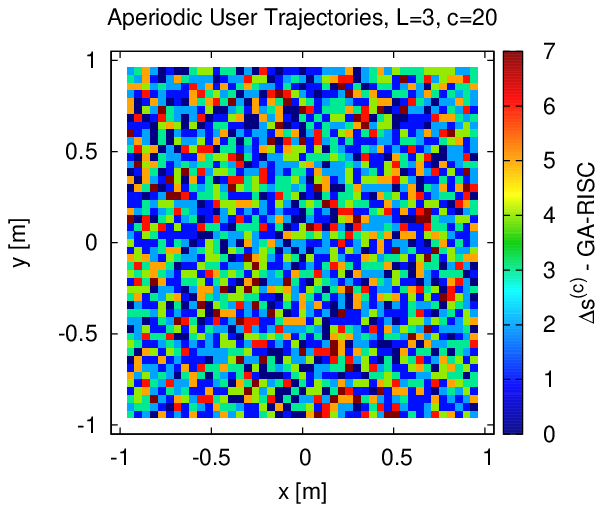}\tabularnewline
(\emph{b})\tabularnewline
\end{tabular}\end{center}

\begin{center}~\vfill\end{center}

\begin{center}\textbf{Fig. 5 - F. Zardi et} \textbf{\emph{al.}}\textbf{,}
\textbf{\emph{{}``}}Memory-Enhanced Dynamic Evolutionary Control...
''\end{center}

\newpage
\begin{center}~\vfill\end{center}

\begin{center}\includegraphics[%
  width=0.95\textwidth]{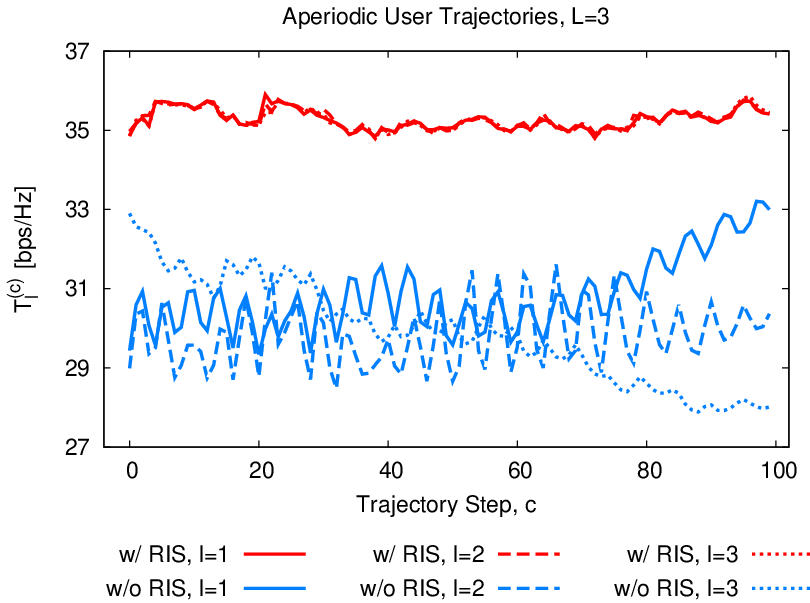}\end{center}

\begin{center}~\vfill\end{center}

\begin{center}\textbf{Fig. 6 - F. Zardi et} \textbf{\emph{al.}}\textbf{,}
\textbf{\emph{{}``}}Memory-Enhanced Dynamic Evolutionary Control...
''\end{center}

\newpage
\begin{center}~\vfill\end{center}

\begin{center}\begin{tabular}{ccc}
&
\emph{w/o~RIS}&
\emph{w/~RIS}\tabularnewline
\begin{sideways}
$\quad\quad l=1$%
\end{sideways}&
\includegraphics[%
  width=0.40\textwidth]{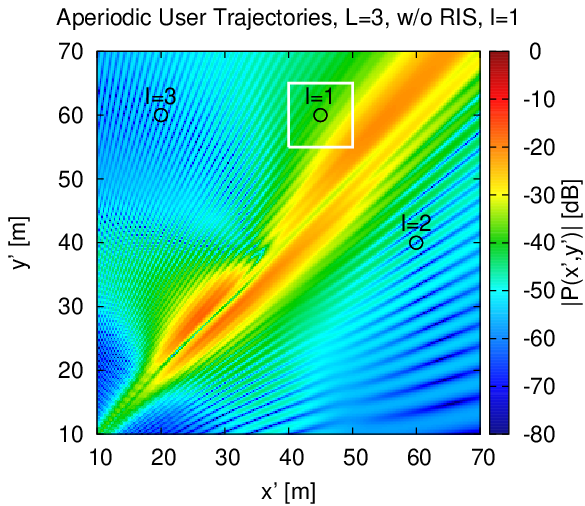}&
\includegraphics[%
  width=0.40\textwidth]{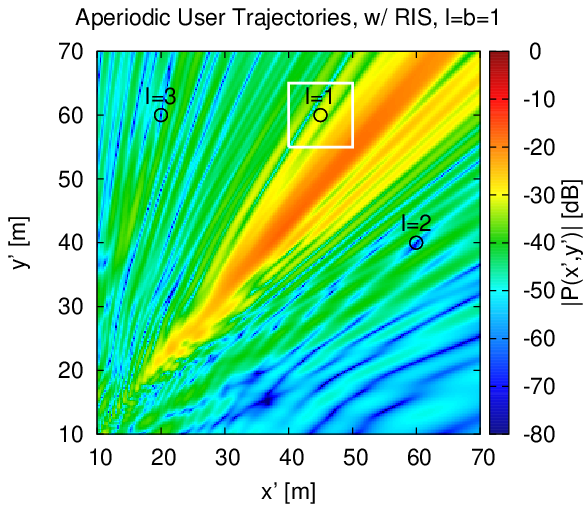}\tabularnewline
&
(\emph{a})&
(\emph{b})\tabularnewline
\begin{sideways}
$\quad\quad l=2$%
\end{sideways}&
\includegraphics[%
  width=0.40\textwidth]{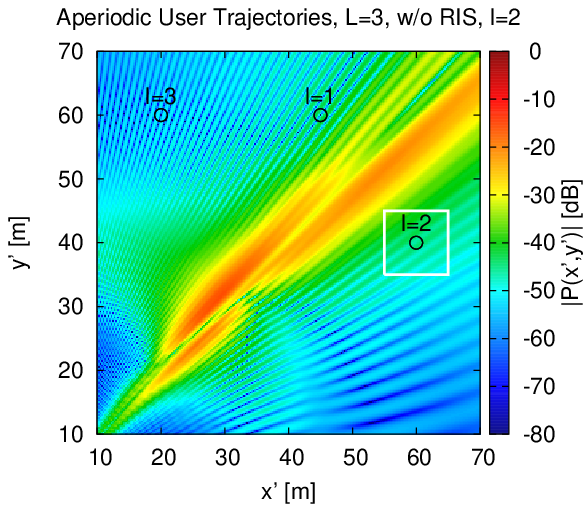}&
\includegraphics[%
  width=0.40\textwidth]{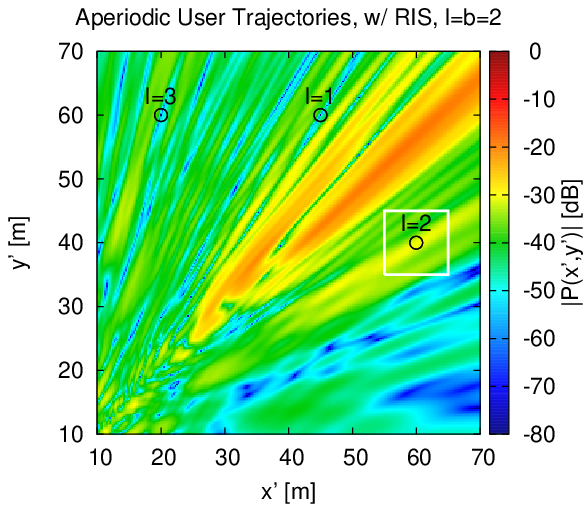}\tabularnewline
&
(\emph{c})&
(\emph{d})\tabularnewline
\begin{sideways}
$\quad\quad l=L=3$%
\end{sideways}&
\includegraphics[%
  width=0.40\textwidth]{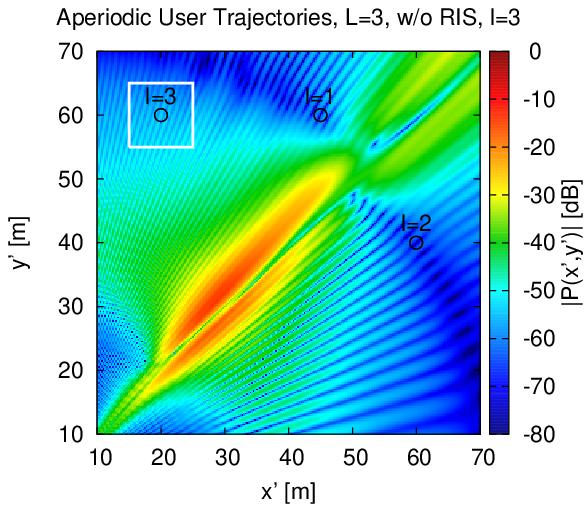}&
\includegraphics[%
  width=0.40\textwidth]{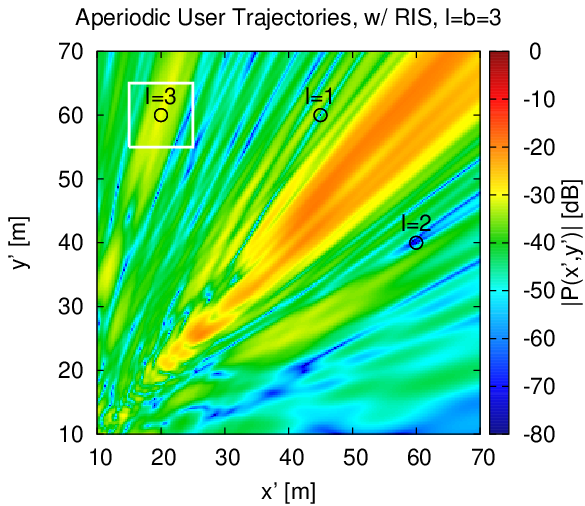}\tabularnewline
&
(\emph{e})&
(\emph{f})\tabularnewline
\end{tabular}\end{center}

\begin{center}~\vfill\end{center}

\begin{center}\textbf{Fig. 7 - F. Zardi et} \textbf{\emph{al.}}\textbf{,}
\textbf{\emph{{}``}}Memory-Enhanced Dynamic Evolutionary Control...
''\end{center}

\newpage
\begin{center}~\vfill\end{center}

\begin{center}\begin{tabular}{c}
\includegraphics[%
  width=0.80\textwidth]{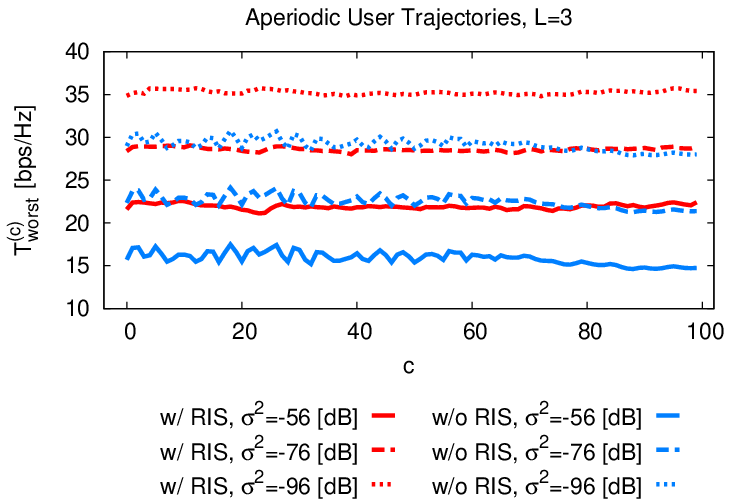}\tabularnewline
(\emph{a})\tabularnewline
\includegraphics[%
  width=0.80\textwidth]{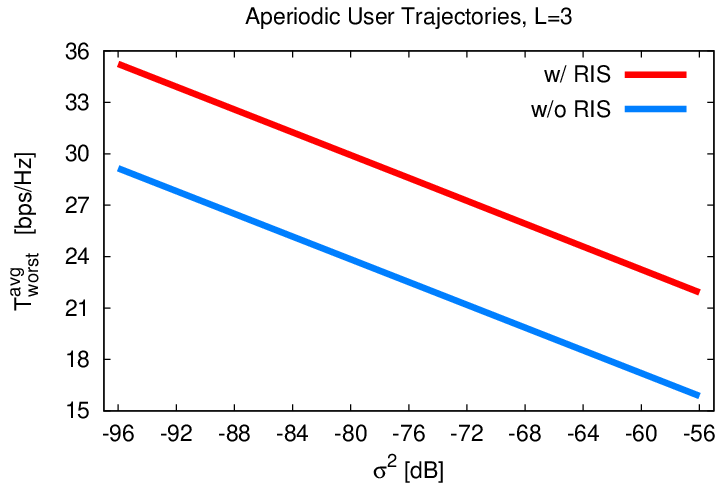}\tabularnewline
(\emph{b})\tabularnewline
\end{tabular}\end{center}

\begin{center}~\vfill\end{center}

\begin{center}\textbf{Fig. 8 - F. Zardi et} \textbf{\emph{al.}}\textbf{,}
\textbf{\emph{{}``}}Memory-Enhanced Dynamic Evolutionary Control...
''\end{center}

\newpage
\begin{center}\begin{tabular}{cc}
\multicolumn{2}{c}{\includegraphics[%
  width=0.75\columnwidth]{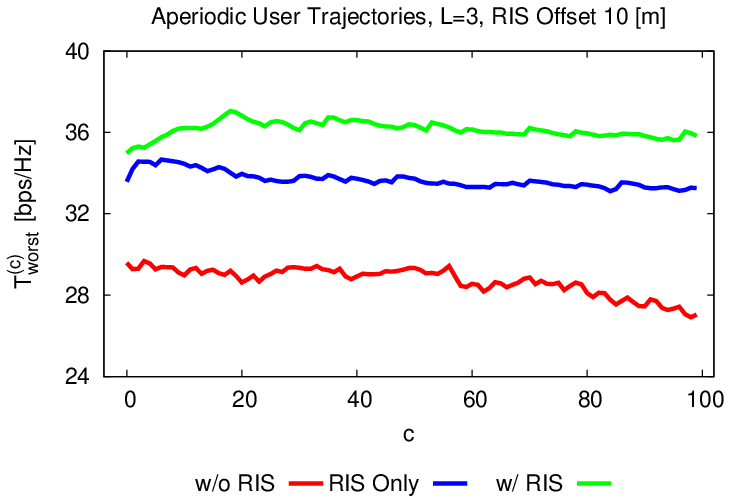}}\tabularnewline
\multicolumn{2}{c}{(\emph{a})}\tabularnewline
\emph{w/o~RIS}&
\emph{RIS only}\tabularnewline
\includegraphics[%
  width=0.40\columnwidth,
  keepaspectratio]{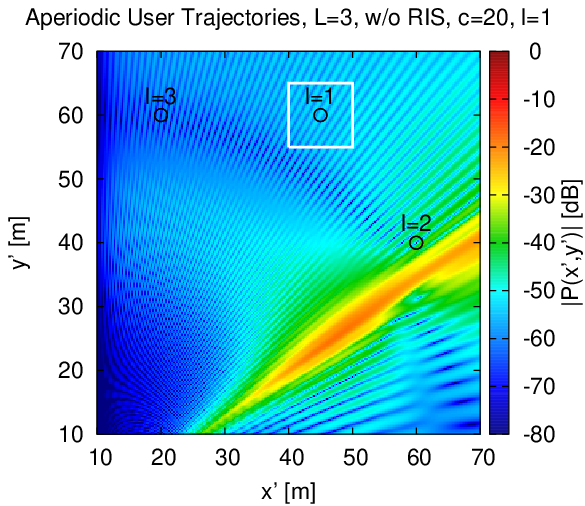}&
\includegraphics[%
  width=0.40\columnwidth,
  keepaspectratio]{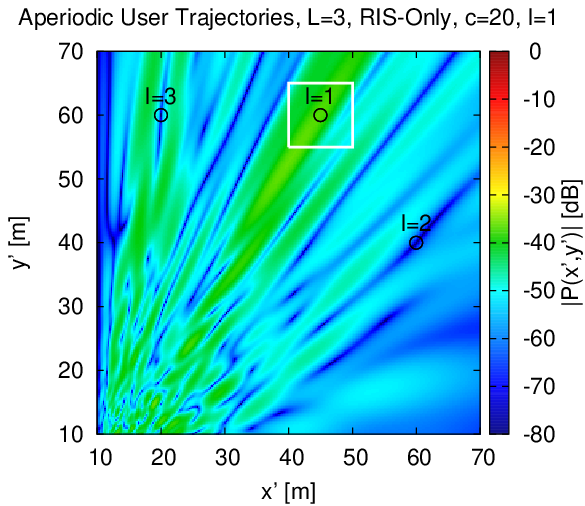}\tabularnewline
(\emph{b})&
(\emph{c})\tabularnewline
\multicolumn{2}{c}{\emph{w~RIS}}\tabularnewline
\multicolumn{2}{c}{\includegraphics[%
  width=0.40\columnwidth,
  keepaspectratio]{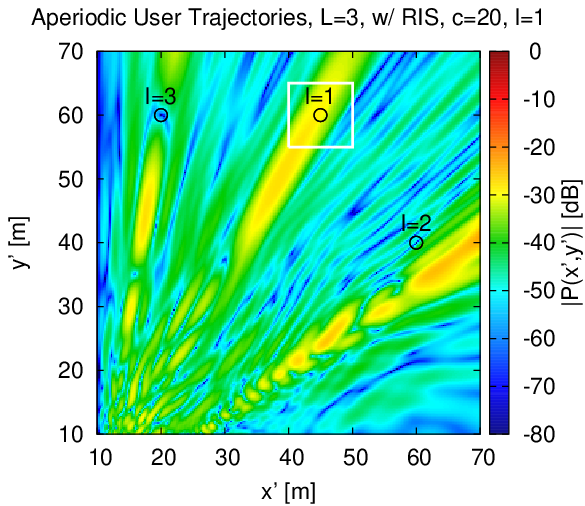}}\tabularnewline
\multicolumn{2}{c}{(\emph{d})}\tabularnewline
\end{tabular}\end{center}

\begin{center}\textbf{Fig. 9 - F. Zardi et} \textbf{\emph{al.}}\textbf{,}
\textbf{\emph{{}``}}Memory-Enhanced Dynamic Evolutionary Control...
''\end{center}

\newpage
\begin{center}~\vfill\end{center}

\begin{center}\begin{tabular}{c}
\includegraphics[%
  width=0.60\columnwidth]{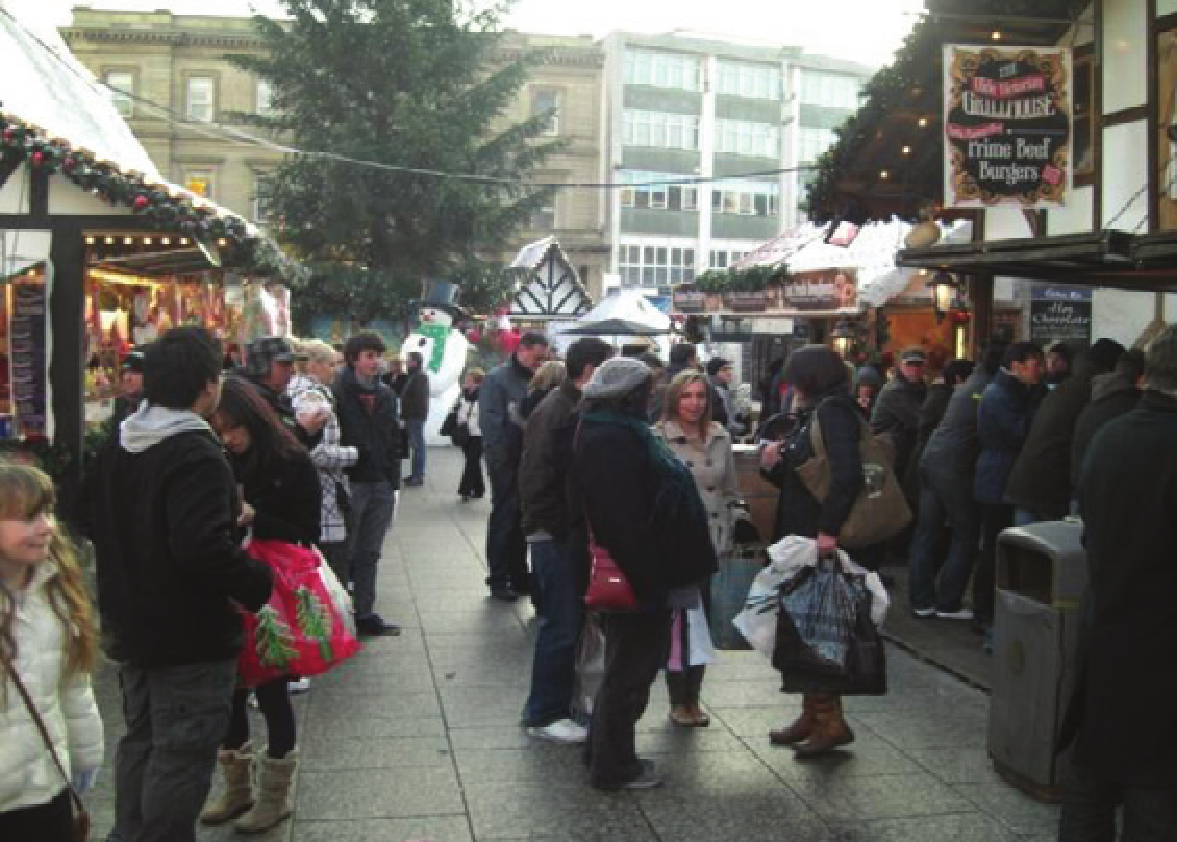}\tabularnewline
(\emph{a})\tabularnewline
\includegraphics[%
  width=0.75\textwidth]{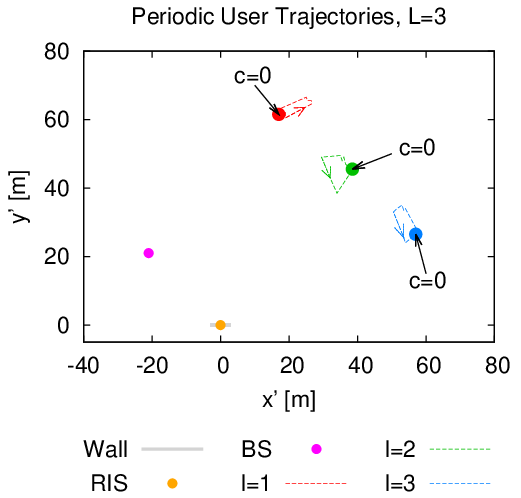}\tabularnewline
(\emph{b})\tabularnewline
\end{tabular}\end{center}

\begin{center}~\vfill\end{center}

\begin{center}\textbf{Fig. 10 - F. Zardi et} \textbf{\emph{al.}}\textbf{,}
\textbf{\emph{{}``}}Memory-Enhanced Dynamic Evolutionary Control...
''\end{center}

\newpage
\begin{center}~\vfill\end{center}

\begin{center}\begin{tabular}{cc}
\multicolumn{2}{c}{\includegraphics[%
  width=0.80\textwidth]{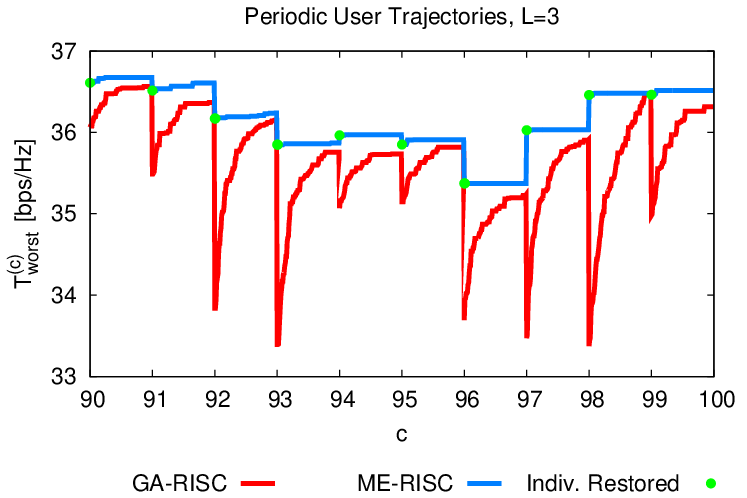}}\tabularnewline
\multicolumn{2}{c}{(\emph{a})}\tabularnewline
\includegraphics[%
  width=0.45\textwidth]{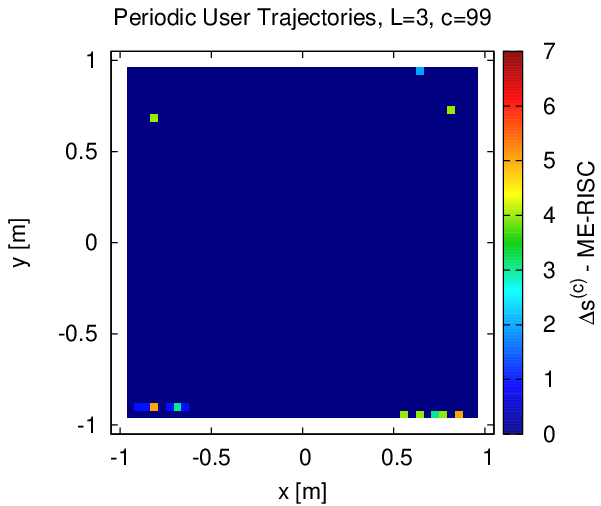}&
\includegraphics[%
  width=0.45\textwidth]{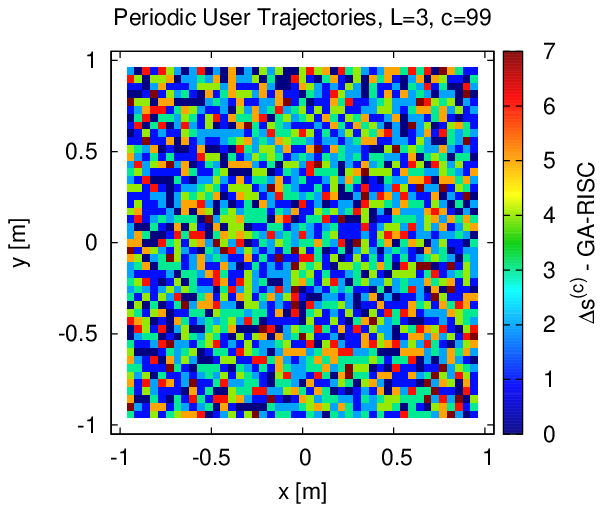}\tabularnewline
(\emph{b})&
(\emph{c})\tabularnewline
\end{tabular}\end{center}

\begin{center}~\vfill\end{center}

\begin{center}\textbf{Fig. 11 - F. Zardi et} \textbf{\emph{al.}}\textbf{,}
\emph{{}``}Memory-Enhanced Dynamic Evolutionary Control... \textbf{''}\end{center}

\newpage
\begin{center}~\vfill\end{center}

\begin{center}\includegraphics[%
  width=0.95\textwidth]{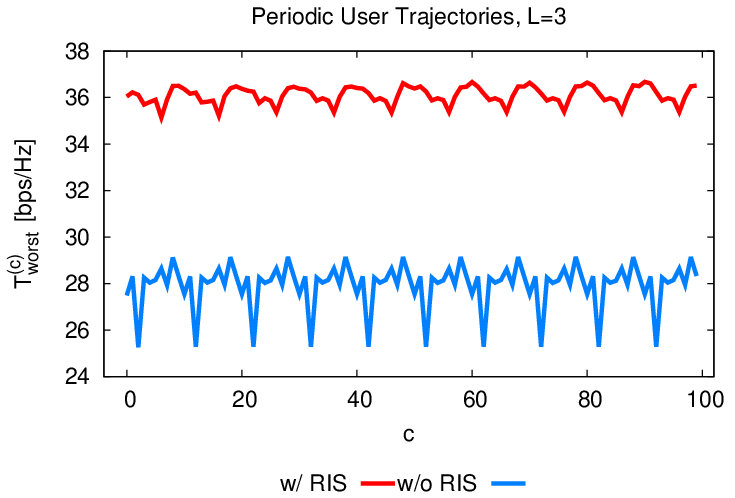}\end{center}

\begin{center}~\vfill\end{center}

\begin{center}\textbf{Fig. 12 - F. Zardi et} \textbf{\emph{al.}}\textbf{,}
\emph{{}``}Memory-Enhanced Dynamic Evolutionary Control... ''\end{center}
\end{document}